\documentclass{JHEP3}
\usepackage{amsmath}
\usepackage{epsfig}
\usepackage{amssymb}
\usepackage{url}
\usepackage{graphics}

\newcommand{\bea}{\begin{eqnarray}}
\newcommand{\eea}{\end{eqnarray}}
\newcommand{\bean}{\begin{eqnarray*}}
	\newcommand{\eean}{\end{eqnarray*}}
\newcommand{\nn}{\nonumber \\}

\def\eref#1{(\ref{#1})}

\def\d{{\rm d}}

\def\a{{\alpha}}

\def\b{{\beta}}

\def\c{{\gamma}}

\def\d{\partial}

\def\eps{\epsilon}

\def\Label#1{\label{#1}%
	\smash{\hbox to0pt{\raise1ex\hbox{\tiny[#1]}\hss}}}

\title{General One-loop Reduction in Generalized Feynman Parametrization Form  }
\date{\today}
\author{ Hongbin Wang
	\footnote{Emails: 21836003@zju.edu.cn } \\
	{\small Zhejiang Institute of Modern Physics, Zhejiang University, Hangzhou, 310027, P. R. China }}
\abstract{ For higher loop computation, one of main topics is to find the effective reduction method. Recently there is an alternative reduction method proposed by Chen in \cite{chen1,chen2}. In this paper,  using the one-loop scalar integrals with  propagators having higher power, we test the power of Chen's new method. More explicitly, with the  
	improved version of the method  we can cancel the dimension shift and terms having unwanted 
	power shifting. Thus the obtained IBP relations
	are much more simpler and can be solved 
	easily. 
	Using this method  we present the explicit examples of bubble, triangle, box and pentagon with one propagators doubled. With these results, we complete our previous computations in \cite{wang} with the missed tadpole coefficients and show the potential of Chen's 
	method for efficient reduction of higher loop
	integrals. 
}


\newpage
\begin{document}
	%
	\section{Introduction}

	To give more precise theoretical prediction of scatting amplitude of a given 
	process, calculation of high loops integrals becomes important. 
	For these calculations, the PV-reduction method \cite{Passarino:1978jh} is one of the
	most used ideas. 
	One way to implement the reduction method is to use 
	Integrating-by-Parts (IBP) relation \cite{smir,ibp1,ibp2}. As one of the most powerful techniques for loop integrals reduction, IBP gives a large number of recurrence relations, and one could get the reduction of the simpler integrals directly by Gauss elimination. However, as the number and power of propagators become higher and higher, the IBP method becomes hard and inefficient. Finding  more efficient reduction methods 
	becomes an important direction. 
	
	Unitarity cut method is one alternative reduction method and
	has been proved to be very useful for one-loop integrals \cite{Bern:1994zx,Bern:1994cg,Britto:2004nc,Cachazo:2004zb,Britto:2005ha,u1,u2,u3,u4,Britto:2009wz,Anastasiou:2006jv,Anastasiou:2006gt}. For physical one-loop process, the power of propagator
	is just one, but if the method is a complete method, it should
	be able to give the reduction of integrals with
	higher power of propagators. Such a situation is not
	just a theoretical curiosity. In fact, it appears in the higher
	loop diagrams as a sub-diagram. Furthermore, although for one-loop integrals the scalar
	basis is natural, in general the choice of basis can be
	different, depending on the physical input. For example,
	for one-loop bubble, the basis with one propagator having
	power two could be useful as part of UT-basis \cite{zhangyanglec,Henn:2014qga}. 
	
	In our previous work \cite{wang}, by combining the trick of differential operators and unitarity cut, we  successfully got the analytical reduction result  of one-loop integrals with high power propagators and gave the coefficients to all the basis except the tadpoles' coefficients. Since the tadpole have only one propagator, the unitarity method could not be used to get the tadpole part.
	To complete our investigation, we want to find the missing 
	tadpole coefficients by some efficient methods.

	Except the unitarity cut method, there are other
	proposals to overcome the difficulty in IBP, by
	using some tricks and other representations of integrals  in recent years, such as Baikov  representation \cite{Baikov:1996rk,Baikov:1996cd}  and Feynman parametrization representation \cite{Bern:1992em,Bern:1993kr} for loop integrals. In recent years, Chen has proposed a new representation for loop integrals \cite{chen1,chen2}. His method is based on  {\sl the generalized Feynman parametrization representation}, i.e., an extra parameter $x_{n+1}$ has been introduced to
	combine the ${\cal U}, {\cal F}$ in the standard Feynman parametrization representation. Such a generalization
	will bring some benefits in deriving the IBP recurrence
	relation, as will shown in this paper.

	As a common feature, the IBP recurrence relation 
	derived using the generalized Feynman parametrization representation
	will naturally have terms in different spacetime dimension. Since we always concern the reduction in a certain dimension $D$, which is usually set to be $4-2\eps$ for the reason of renormalization, we want to cancel these terms in different dimension. This is usually not an easy work. 
	In \cite{Kosower} Gluza, Kajda and Kosower have shown how to avoid the change of power of propagators in the standard momentum space. Larsen and Zhang  have considered the Baikov representation and showed how to eliminate both dimension shifting and the change
	of power of propagators \cite{zhang1,zhang2,Larsen:2015ped,Larsen:2016tdk,Zhang:2016kfo,Jiang:2017phk}. These methods require the solution of syzygy equations, which is not  easy to figure out in general. In Chen's second paper \cite{chen2}, he proposed a new technique to simplifying the recurrence relation based on the non-commutative algebra.

	Motivated by above discussion and preparing Chen's method  for the  high-loop computations, in this paper, we will
	use the Chen's method to find the missing tadpole coefficients
	in our previous work. Furthermore, we will use 
	the idea to remove terms with dimensional shifting in the
	derived IBP relation to give a simpler reduction method with the analytic results written by the elements of the coefficients matrix $\hat A$.
	
	The plan of the paper is following. 
	In section 2, we have  reviewed the Chen's new method and illustrated  with a simple example in the section 2.1. 
	In the example, the integrals in different dimension will naturally emerge. We discussed the physical meaning of the boundary terms, which contributes to the sub-topologies. To cancel the dimension in the parametrization form and simplify the IBP relation, in the section 2.2 we proposed a new trick by adding free auxiliary parameters based on the fact that the  $F$ in the integrand is a homogeneous function of $x_{i}$ with degree $L+1$.  By our trick, we successfully canceled the dimension shift and dropped the terms that we do not concern to a certain extent, and give a simplified IBP relation in which all the integrals are in the certain dimension $D$ and integrals except the target have lower total power of  propagators. We gave our analytic result by the determinant of the  cofactor of the matrix $\hat A$, which is completely  determined by the graph.   In section 3,  combined with our trick, we calculated the triangle $I_3(1,1,2)$, box  $I_4(1,1,1,2)$, and pentagon $I_5(1,1,1,1,2)$ in the parametric form proposed by Chen , and gave the analytic result of all the coefficients to the master basis, especially the tadpole parts as the complement of our previous work. 
	\section{Reduction method  in parametric form by Chen}
	
	In this section, we will introduce a new reduction method proposed by Chen in \cite{chen1}.
	The general form of loop integral is given by
	\bea
	I[N(l)](k)&=&\int d^{D}l_1d^{D}l_2\cdots d^{D}l_{L}\frac{N(l)}{D_1^{k_1}D_{2}^{k_2}D_3^{k_3}\cdots D_{n}^{k_n}}~~~\label{Wang-2-1}
	\eea
	where for simplicity, we have denoted $l=(l_1,l_2,l_3,\cdots ,l_L)$ and
	$k=(k_1,k_2,k_3,\cdots ,k_n)$. Since in this paper, we consider only the scalar's integrals with  $N(l)=1$,  let us label
	\bea
	I(L;\lambda_1+1,\cdots ,\lambda_n+1)&=&\int d^Dl_1\cdots d^Dl_L\frac{1}{D_1^{\lambda_1+1}\cdots D_{n}^{\lambda_n+1}}~~~\label{Wang-2-2}
	\eea
	By the procedure of Feynman parametrization,
	\bea
	\sum_{i}^{L}\a_iD_i&=&\sum_{i,j}^{L}A_{ij}l_i\cdot l_j+2\sum_{i=1}^{L}B_i\cdot l_i+C~~~\label{Wang-2-5}
	\eea
	thus  the loop integrals can be done as
	\bea
	\int d^Dl_1\cdots d^Dl_L e^{i(\sum\a_iD_i)}&=&e^{i\pi L(1-\frac{D}{2})/2}\pi^{LD/2}(Det~A)^{-\frac{D}{2}}~e^{i(C-\sum A_{ij}^{-1}B_i\cdot B_j)}~~~\label{Wang-2-6}
	\eea
	Defining $U(\a)=Det~A$, and
	$C-\sum A_{ij}^{-1}B_i\cdot B_j\equiv \frac{V(\a)}{U(\a)}-\sum m_i^2\a_i$~~\footnote{The relation has been verified in many places based on the method in graph theory}, one can see that $U(\a)$ is a homogeneous function of $\a_i$ with degree $L$, while the $V(\a)$ is a homogeneous function of $\a_i$ with degree $L+1$, and  the loop integral becomes to
	\bea
	I(L;\lambda_1+1,\cdots ,\lambda_n+1)&=&\frac{e^{-\sum \frac{\lambda_i+1}{2}i\pi}}{\Pi_{i=1}^n\Gamma(\lambda_i+1)}e^{i\pi L(1-\frac{D}{2})/2}\pi^{LD/2}\nn
	&&\times\int d\a_1\cdots d\a_n U(\a)^{-\frac{D}{2}}e^{i[V(\a)/U(\a)-\sum m_i^2\a_i]}\a_1^{\lambda_1}\cdots \a_n^{\lambda_n}~~~\label{IL}
	\eea
	To derive the parametric form suggested by Chen, we do the following.
	Using the $\a$-representation of general propagators,
	\bea
	\frac{1}{(l^2-m^2)^{\lambda+1}}&=&\frac{e^{-\frac{\lambda+1}{2}i\pi}}{\Gamma(\lambda+1)}\int _0^{\infty}d\a e^{i\a(l^2-m^2)}\a^{\lambda},~~~~~~
	Im\{l^2-m^2\}>0~~~\label{Wang-2-3}
	\eea
	where the "$i\eps$" has been neglected, we get
	\bea
	I(L;\lambda_1+1,\cdots ,\lambda_n+1)&=&\frac{e^{-\sum_i^n \frac{\lambda_i+1}{2}i\pi}}{\Pi_{i=1}^{n}\Gamma(\lambda_i+1)}\int d^Dl_1\cdots d^Dl_L\int _0^{\infty}d\a_1\cdots d\a_n e^{i\sum_{i=1}^{n}\a_iD_i}\a_1^{\lambda_1}\cdots \a_n^{\lambda_n}~~~\label{Wang-2-4}
	\eea
	To go further, we change the integral variables as $\a_i=\eta x_i$.
	Since there are totally $n$ independent variables, we must put another constraint condition. In general, we could let
	\bea
	\sum_{i\in S(1,2,3,\cdots n)} x_i=1~~~\label{Wang-2-7}
	\eea
	where $S$ is an arbitrary  non-trivial subset of $\{1,2,3,\cdots n\}$.
	After carrying out the integration over  $\eta$,  the second line of eq.\eref{IL} becomes to
	\bea
	&&(-i)^{(n+\lambda-\frac{DL}{2})}\Gamma(n+ \lambda-\frac{DL}{2})\times\int dx_1\cdots dx_n\delta (\sum_{j\in S}x_j-1) \frac{U(x)^{n+\lambda-\frac{D}{2}(L+1)}}{[-V(x)+U(x)\sum m_i^2x_i]^{n+\lambda-\frac{DL}{2}}}x_1^{\lambda_1}\cdots x_n^{\lambda_n}\nn
	&=&(-i)^{n+ \lambda-\frac{DL}{2}}\Gamma(n+\lambda-\frac{DL}{2})\int dx_1\cdots dx_n\delta (\sum_{j\in S} x_j-1)U^{\lambda_u}f^{\lambda_f}x_1^{\lambda_1}\cdots x_n^{\lambda_n}~~~\label{1.18}
	\eea
	where
	\bea
	U(x) & = & \eta^{-L} U(\a)= \eta^{-L} U(\eta x_i),~~~~V(x)  =  \eta^{-L-1} V(\a)= \eta^{-L-1} V(\eta x),~~~~f(x)=-V(x)+U(x)\sum m_i^2x_i\nn
	\lambda&=&\sum_{i=1}^{n} \lambda_i,~~~~
	\lambda_u=n+\lambda-\frac{D}{2}(L+1),~~~~
	\lambda_f=-n-\lambda+\frac{DL}{2}~~~\label{Wang-2-8}
	\eea
	Finally we by  Mellin transformation\footnote{Different from the traditional Feynman parametrization, here we should add a new auxiliary parameter $x_{n+1}$ to transform the integral into a symmetric form.1}
	\bea
	A^{\lambda_1}B^{\lambda_2}&=&\frac{\Gamma(-\lambda_1-\lambda_2)}{\Gamma(-\lambda_1)\Gamma(-\lambda_2)}\int _0^{\infty} dx  (A+Bx)^{\lambda_1+\lambda_2}x^{-\lambda_2-1}~~~\label{Wang-2-9}
	\eea
	we could  write  the \eref{1.18} as
	\bea
	&&(-i)^{n+\lambda-\frac{DL}{2}}\Gamma(n+\lambda-\frac{DL}{2})\frac{\Gamma(-\lambda_u-\lambda_f)}{\Gamma(-\lambda_u)\Gamma(-\lambda_f)}\int dx_1 \cdots dx_n \delta(\sum_{j\in S}x_j-1)\int _0^\infty dx_{n+1}\nn%
	&&~~~~~~~~~~~~~~~~~~~~~~~~~~\times (Ux_{n+1}+f)^{\lambda_u+\lambda_f}x_{n+1}^{-\lambda_u-1}x_1^{\lambda_1} \cdots x_n^{\lambda_n}\nn%
	&\equiv& (-i)^{n+\lambda-\frac{DL}{2}}\frac{\Gamma(n+\lambda-\frac{DL}{2})\Gamma(-\lambda_u-\lambda_f)}{\Gamma(-\lambda_u)\Gamma(-\lambda_f)}\int d\Pi^{(n+1)}F^{\lambda_0}x_1^{\lambda_1}\cdots x_n^{\lambda_n} x_{n+1}^{\lambda_{n+1}}\nn%
	&\equiv & (-i)^{n+\lambda-\frac{DL}{2}}\frac{\Gamma(n+\lambda-\frac{DL}{2})\Gamma(-\lambda_u-\lambda_f)}{\Gamma(-\lambda_u)\Gamma(-\lambda_f)} i_{\lambda_0;\lambda_1,\cdots \lambda_n}~~~\label{Wang-2-10}
	\eea
	where
	\bea
	d\Pi^{(n+1)}&=&dx_1\cdots dx_{n+1}\delta (\sum_{j\in S} x_j-1),~~~~
	F=Ux_{n+1}+f,~~~~~\lambda =\sum_{i=1}^n \lambda_i,\nn
	\lambda_0&=&\lambda_u+\lambda_f=-\frac{D}{2},~~~~
	\lambda_{n+1}=-\lambda_u-1=\frac{D}{2}(L+1)-\lambda-1-n~~~\label{Wang-2-11}
	\eea
	Putting all together, we now finally get the parametric form of scalar loop integrals \eref{IL},
	\bea
	I(L;\lambda_1+1,\cdots ,\lambda_n+1)&=&(-1)^{n+\lambda}i^{L} \pi^{\frac{LD}{2}}\frac{\Gamma(-\lambda_0)}{\Pi_{i=1}^{n+1}\Gamma(\lambda_i+1)} i_{\lambda_0;\lambda_1,\cdots \lambda_n}~~~\label{Wang-2-12}
	\eea
	\subsection{The IBP identity in parametric represent}
	%
	
	The parametric form of \eref{Wang-2-12} is the starting point of Chen's proposal. The IBP relations in this form is given by\footnote{In some sense, the parametric form can be considered as the {\bf generalized Feynman parametrization form}. Thus the IBP relation \eref{ibp1} could be
		called the IBP relation in the  generalized Feynman parametrization form. }\footnote{The IBP relation requires the term in the bracket of the first term to be degree $(-n)$, which can be obtained by multiplying any monomial of degree one. Here in \eref{ibp1} we have multiplied 
		$x_{n+1}$ by our experiences from later examples, but one can make other choices. }
	\bea
	\int d\Pi^{(n+1)}\frac{\d }{\d x_i}\Big\{F^{\lambda_0}x_1^{\lambda_1}\cdots x_n^{\lambda_n}x_{n+1}^{\lambda_{n+1}+1}\Big\}+\delta_{\lambda_i,0}\int d\Pi^{(n)} \Big\{F^{\lambda_0}x_1^{\lambda_1}\cdots x_n^{\lambda_n}x_{n+1}^{\lambda_{n+1}+1}\Big\}\Big|_{x_i=0}=0~~\label{ibp1}
	\eea
	where $i=1,...,n+1$ and the $d\Pi^{(n)}$ in the second term is
	\bea
	d\Pi^{(n)}&=&dx_1\cdots \hat{dx_i}\cdots  dx_ndx_{n+1}\delta(\sum_{j\in S}x_j-1)~~~\label{Wang-2-13}
	\eea
	The second term in \eref{ibp1} contributes to a boundary term which leads to the sub-topologies to the former term. 

	To illustrate the IBP relation \eref{ibp1}, we present the  reduction of $I_2(1,2)$ as an example.
	The general form of one-loop bubble integrals is given by
	\bea
	I_2(m+1,n+1)&=&\int \frac{d^{D}l}{(l^2-m_1^2)^{m+1}((l-p_1)^2-m_2^2)^{n+1}}~~~\label{Wang-bub-2-1}
	\eea
	and the corresponding  parametric form  is  (in this article we ignore the former factor $\pi^{\frac{LD}{2}}$)
	\bea
	I_2(m+1,n+1)&=&i(-1)^{m+n+2}\frac{\Gamma(\frac{D}{2})}{\Gamma(m+1)\Gamma(n+1)\Gamma(D-2-m-n)}\int d\Pi^{(3)}F^{\lambda_0}x_1^mx_2^nx_3^{\lambda_3}~~~\label{Wang-bub-2-2}
	\eea
	where
	\bea
	F&=&(x_1+x_2)(m_1^2x_1+m_2^2x_2+x_3)-p_1^2x_1x_2~~~\label{Wang-bub-2-3}
	\eea
	and
	\bea
	i_{\lambda_0;m,n}&=&\int d\Pi^{(3)}F^{\lambda_0}x_1^mx_2^nx_3^{\lambda_3}~~~\label{Wang-bub-2-4}
	\eea
	with $\lambda_0=-\frac{D}{2}$ and
	$\lambda_3=-3-m-n-2\lambda_0$.
	Using the eq.\eref{ibp1},  we could get three IBP recurrence relations. Taking $\frac{\d}{\d x_1}$ first, the first term in \eref{ibp1} gives
	\bea
	\lambda_0i_{\lambda_0-1;m,n}+2m_1^2\lambda_0 i_{\lambda_0-1;m+1,n}+\Delta\lambda_0i_{\lambda_0-1;m,n+1}~~~\label{Wang-bub-2-5}
	\eea
	where $\Delta=m_1^2+m_2^2-p_1^2$. The second term gives
	\bea
	& & \delta_{m,0}\int d\Pi^{(2)}(x_3+m_2^2x_2)^{\lambda_0}x_2^{n+\lambda_0}x_3^{-2-n-2\lambda_0}=\delta_{m,0}i_{\lambda_0;-1,n}
	~~~\label{Wang-bub-2-6}
	\eea
	Here we need to explain the notation $i_{\lambda_0;-1,n}$. From the middle expression of \eref{Wang-bub-2-6}, we see that 
	it is the parametric form of tadpole $\int \frac{d^{D}l}{(l^2-m_2^2)^{n+1}}$. To emphasize its origin, i.e., coming from bubble by removing 
	the first propagator, we extend the definition of $i_{\lambda_0;\lambda_1,...,\lambda_n}$ given in \eref{Wang-2-10} by 
	setting $\lambda_1=-1$~~\footnote{Same notation has also been used in \cite{chen2} (see Eq. (2.5a)).}.  Using the extended notation,  we got the  first IBP relation
	\bea
	\lambda_0i_{\lambda_0-1;m,n}+2m_1^2\lambda_0 i_{\lambda_0-1;m+1,n}+\Delta\lambda_0i_{\lambda_0-1;m,n+1}+\delta_{m,0}i_{\lambda_0;-1,n}&=&0
	~~~\label{Wang-bub-2-7}%
	\eea
	When we set $m=n=0$ in \eref{Wang-bub-2-7}, it reads
	\bea
	\lambda_0i_{\lambda_0-1;0,0}+2m_1^2\lambda_0i_{\lambda_0-1;1,0}+
	\Delta\lambda_0i_{\lambda_0-1;0,1}+i_{\lambda_0;-1,0}&=&0~~~\label{Wang-bub-2-8}
	\eea
	Similarly, we could take the differential $\frac{\d }{\d x_2}$ and get the second IBP relation
	\bea
	\lambda_0i_{\lambda_0-1;0,0}+\Delta\lambda_0 i_{\lambda_0-1;1,0}+2m_2^2\lambda_0i_{\lambda_0-1;0,1}+i_{\lambda_0;0,-1}&=&0~~~\label{Wang-bub-2-9}
	\eea
	Naively, we should solve $i_{\lambda_0;0,1}$ by $i_{\lambda_0;0,0}$ from \eref{Wang-bub-2-8} and \eref{Wang-bub-2-9}. However, for bubble part, we have $\lambda_0-1$ instead of $\lambda_0$.
	This one could be fixed by rewriting $\lambda_0\to \lambda_0+1$ since $\lambda_0$ is a free parameter. However, the boundary tadpole part $i_{\lambda_0;0,-1}$ will become $i_{\lambda_0+1;0,-1}$, i.e., having the dimensional shifting, which is
	a common feature in the parametric IBP relation.

	To deal with it, using the parametric form of tadpoles
	\bea
	i_{\lambda_0;m,-1}&=&\int d\Pi^{(2)}(x_1x_3+m_1^2x_1^2)^{\lambda_0}x_1^mx_3^{-2-m-2\lambda_0}
	~~~\label{Wang-bub-2-10}%
	\eea
	and taking the $\frac{\d }{\d x_1}$ and $\frac{\d }{\d x_3}$, we could get   two  IBP relations
	\bea
	\lambda_0 i_{\lambda_0-1;m,-1}+2m_1^2\lambda_0 i_{\lambda_0-1;m+1,-1}+m i_{\lambda_0,m-1,-1}&=&0\nn
	\lambda_0 i_{\lambda_0-1;m+1,-1}+(-1-m-2\lambda_0)i_{\lambda_0;m,-1}&=&0~~~\label{Wang-bub-2-11}
	\eea
	from which we solve
	\bea
	i_{\lambda_0;0,-1}&=&\frac{-\lambda_0}{2m_1^2(2\lambda_0+1)}i_{\lambda_0-1;0,-1},~~~~
	i_{\lambda_0;-1,0}=\frac{-\lambda_0}{2m_2^2(2\lambda_0+1)}i_{\lambda_0-1;-1,0}~~~\label{Wang-bub-2-12}
	\eea
	Putting \eref{Wang-bub-2-12} to \eref{Wang-bub-2-8} and \eref{Wang-bub-2-9}, we can solve the $i_{\lambda_0-1;0,1}$. After  shifting $\lambda_0\to \lambda_0+1$, we finally get
	\bea
	i_{\lambda_0;0,1}=\frac{2m_1^2-\Delta}{\Delta^2-4m_1^2m_2^2}i_{\lambda_0;0,0}+\frac{-1}
	{(2\lambda_0+3)(\Delta^2-4m_1^2m_2^2)}i_{\lambda_0;0,-1}+\frac{\Delta}{2m_2^2(2\lambda_0+3)
		(\Delta^2-4m_1^2m_2^2)}i_{\lambda_0;-1,0}~~~~~\label{Wang-bub-2-13}
	\eea
	Translating back to scalar basis, we get the reduction of $I_2(1,2)$ as
	\bea
	I_2(1,2)&=&c_{2\to2}I_2(1,1)+c_{2\to1\bar2}I_2(1,0)+c_{2\to1;\bar1}I_2(0,1)~~~\label{Wang-bub-2-14}
	\eea
	with the coefficients
	\bea
	c_{2\to2}&=&\frac{(D-3)(\Delta-2m_1^2)}{\Delta^2-4m_1^2m_2^2},~~~
	c_{2\to1;\bar2}=\frac{D-2}{\Delta^2-4m_1^2m_2^2},~~~
	c_{2\to1;\bar1}=\frac{(D-2)\Delta}{2m_2^2(4m_1^2m_2^2-\Delta^2)}~~~\label{Wang-bub-2-15}
	\eea
	The result is confirmed with the FIRE6\cite{Smirnov:2019qkx,Lee:2013mka}.
	\subsection{Improvement of parametric IBP}

	As we have seen from the previous subsection, the IBP relation given in \eref{ibp1} will contain the integrals with dimension shift, which makes the reduction program  a bit troublesome. We would like a recurrence relation without dimension shift. As we reviewed in the introduction  there are several references dealt with this or related problems. Based on these work, an improved version of IBP relation 
	has been given in \cite{chen2} (see Eq.(2.12), (2.13) ). All these methods require the solution of syzygy equations, which is not an easy task in general. However, for our one-loop integrals, the function $F(x)$ is a homogeneous function of $x_i$ with degree two\footnote{Note the $F(x)$ is a homogeneous function of degree $L+1$ where $L$ is the number of loops.}. This good property makes  the related syzygy equations simple, which can be solved straightly\footnote{In general, this trick could be extended to high loops to avoid the troublesome calculation of syzygy equations. }. In this paper, we will develop a direct algorithm to write down IBP relations without the dimension shift and the terms having unwanted higher power  of  propagators. 

	In the generalized parametric representation, our improved IBP relation is to multiply a degree zero coefficient $z_i$, for example, $z_i=x_1^{\a}x_2^{\b}x_3^{-\a-\b}$, in \eref{ibp1}. Since the degree of the new integrand does not change, the IBP identity still holds.
	Summing them together we get\footnote{Note the summation of $i $ is form $1$ to $n+1$, where we have included the auxiliary parameter $x_{n+1}$, which is an apparent different from the tradition Feynman parametrization.}
	\bea
	\sum_{i=1}^{n+1}\int d\Pi^{(n+1)}\frac{\d}{\d x_i}\Big\{z_iF^{\lambda_0}x_1^{\lambda_1}x_2^{\lambda_2}\cdots x_{n+1}^{\lambda_{n+1}+1}\Big\}+\sum_{i=1}^{n+1}\delta_{\lambda_i,0}\int d\Pi^{(n)}z_i F^{\lambda_0}x_1^{\lambda_1}\cdots x_n^{\lambda_n}x_{n+1}^{\lambda_{n+1}}|_{x_i=0}&=&0~~~\label{bubb}
	\eea
	Since the second boundary term involve integrals with sub-topologies,  we focus on the first term. Expanding it, we got
	\bea
	\int d\Pi^{(n+1)}\Big[\sum_{i=1}^{n+1}\Big(\frac{\d z_i}{\d x_i}+\lambda_0\frac{z_i\frac{\d F}{\d x_i}}{F}+\lambda_i\frac{z_i}{x_i}\Big)+\frac{z_{n+1}}{x_{n+1}}\Big]F^{\lambda_0}x_1^{\lambda_i}x_2^{\lambda_2}\cdots x_n^{\lambda_n}x_{n+1}^{\lambda_{n+1}+1}
	\eea
	From  \eref{Wang-2-11}, one can see the power $\lambda_0$ of $F$ is related to dimension. 
	To cancel the dimension shift, we need to choose the proper coefficients $z_i$ so that the $\sum_{i=1}^{n+1}z_i\frac{\d F}{\d x_i}$ is a multiple of  the function $F$, i.e.,
	\bea
	\sum_{i=1}^{n+1}z_i\frac{\d F}{\d x_i}+BF&=&0~.~~\label{BBB}
	\eea
	Since  coefficients $z_i$ are not polynomials, \eref{BBB} is not the "normal sygyzy equation" and  one can not directly use the technique developed for polynomial ring. In  \cite{chen2}, Chen developed a method based on the lift and down operators. Here for the  one loop integrals, we can
	solve it directly with some free auxiliary parameters, as we will show shortly. When putting back solutions to the IBP recurrence relation, we could  choose these free parameters to cancel both the dimension shift and unwanted terms with higher power of propagators, which leads to  a simpler recurrence relation.

	Now let us explain the idea in details. Note that in one loop case, the homogeneous function $F$ is a degree two function of $x_i$, so we can write 
	\bea F= {1\over 2} A_{ij} x_i x_j\eea
	where $A$ is the symmetric matrix\footnote{In general it is not necessary to make  $\hat A$ be symmetry matrix, and this is just one choice. But for the simplification of the following calculation, since we will later set an antisymmetric matrix $\hat K_{A}$, it is convenient to make the convention to set $\hat A$ be symmetry matrix.  }. Thus we have 
	\bea f_i\equiv \frac{\d F}{\d x_i},~~~~~\hat f&=&\hat A \hat x,~~~\hat f\equiv \left[\begin{array}{c}
		f_1\nn
		f_2\nn
		\vdots\nn
		f_n\nn
		f_{n+1}
	\end{array}\right],~~
	\hat x\equiv \left[\begin{array}{c}
		x_1\nn
		x_2\nn
		\vdots\nn
		x_n\nn
		x_{n+1}
	\end{array}\right],
	\eea
	Solving $\hat x=\hat A^{-1} \hat f$, we have 
	\bea F={1\over 2} \hat x^T A \hat x={1\over 2}  \hat f^T (\hat A^{-1})^T \hat A \hat A^{-1} \hat f= {1\over 2}  \hat f^T (\hat A^{-1})^T \hat f\equiv \hat f^{T} \hat K \hat f,~~~~~K={1\over 2}A^{-1} ~~~\label{K}\eea
	where the coefficients' matrix $\hat K$ is a real symmetry matrix. In fact we can do more.   Using the trick that
	\bea
	0&=&\hat f^{T} \hat K_{A} \hat f~~~\label{KA}
	\eea
	with any antisymmetric matrix $K_{A}$, we could add  \eref{KA} to \eref{K} to get a more general form
	\bea
	F&=&\hat f^{T}\hat K\hat f+\hat f^T\hat K_A \hat f
	=\hat f^T (\hat K+\hat K_A)\hat f
	\equiv\hat f^T \hat R \hat f
	=\hat f^T\hat R\hat A\hat x
	\equiv \hat f^T\hat Q\hat x,~~~\hat Q\equiv {1\over 2}\hat I+ \hat K_A\hat  A~~~\label{49}
	\eea
	Noticing that because the arbitrary matrix $\hat K_A$ of rank $n+1$, there are $\frac{n(n+1)}{2}$ free independent parameters, $a_{1},\cdots, a_{\frac{n(n+1)}{2}}$ in the matrix $\hat Q$ in 
	\eref{49}.

	Now putting \eref{49} back to \eref{BBB}, we could solve $\hat z$ as
	\bea \hat f^T \hat z+ B \hat f^T\hat Q\hat x=0,~~~\Longrightarrow \hat z= -B\hat Q\hat x~~~\label{z-sol}\eea
	Noticing that since $z$ is degree zero, we should have $B$ homogenous function of degree $-1$. In our article, we choose  $B=\frac{1}{x_{n+1}}$. The choice of $z$ given by \eref{z-sol} will guarantee to remove the dimension shift in the IBP relation. Furthermore,  
	by choosing particular value of  these free parameters of $\hat Q$, we could cancel some unwanted  terms. In the later computations, we will give some examples to illustrate this trick.
	%
	%
	%
	%
	\section{Reduction of one-loop integrals}
	As we have mentioned in the introduction, one motivation of the paper is to complete the reduction
	of scalar basis with general powers. Using the unitarity cut method in \cite{wang}, we are able to find
	reduction coefficients of all basis, except the tadpole. In this section, we will use the improved 
	IBP relation \eref{bubb} to find the tadpole coefficients as well as other coefficients.
	
	\subsection{The bubble's case}
	%
	Let us start from the bubble topology. Although we have done it already in \eref{Wang-bub-2-14}, 
	here we will redo it using the improved IBP relation \eref{bubb}.
	The parametric form of bubble is given by  \eref{Wang-bub-2-2}, \eref{Wang-bub-2-3} and \eref{Wang-bub-2-4}.
	Using our label, we have 
	\bea
	\hat f&=&\hat A \hat x,~~~~~\hat A=\left[\begin{array}{ccc}
		2m_1^2&\Delta&1\nn
		\Delta&2m_2^2&1\nn
		1&1&0
	\end{array}\right]\nn ~~~~\label{Wang-bub-1}
	\eea
	and
	\bea
	F&=&\hat f^T \hat K\hat f,~~~~~\hat K
	=\left[\begin{array}{ccc}
		\frac{1}{4p_1^2}&-\frac{1}{4p_1^2}&\frac{-m_1^2+m_2^2+p_1^2}{4p_1^2}\nn
		-\frac{1}{4p_1^2}&\frac{1}{4p_1^2}&\frac{m_1^2-m_2^2+p_1^2}{4p_1^2}\nn
		\frac{-m_1^2+m_2^2+p_1^2}{4p_1^2}&\frac{m_1^2-m_2^2+p_1^2}{4p_1^2}&\frac{\Delta^2-4m_1^2m_2^2}{4p_1^2}
	\end{array}\right]\nn ~~~~\label{Wang-bub-2}
	\eea
	Adding the antisymmetric matrix $K_A$, we have 
	\bea
	\hat K_A&=&\left[\begin{array}{ccc}
		0&a_1&a_2\nn
		-a_1&0&a_3\nn
		-a_2&-a_3&0\nn
	\end{array}\right],~~
	\hat Q=\left[\begin{array}{ccc}
		\frac{1+2a_2+2a_1m_1^2+2a_1m_2^2-2a_1p_1^2}{2}&a_2+2a_1m_2^2&a_1\nn
		a_3-2a_1m_1^2&\frac{1+2a_3-2a_1m_1^2-2a_1m_2^2+2a_1p_1^2}{2}&-a_1\nn
		-2a_2m_1^2-a_3(m_1^2+m_2^2-p_1^2)&-2a_3m_2^2-a_2(m_1^2+m_2^2-p_1^2)&\frac{1-2a_2-2a_3}{2}
	\end{array}\right]\nn ~~~~\label{Wang-bub-3}
	\eea
	%
	\subsubsection{Deriving the recurrence relation}
	Taking $B={-1\over x_3}$ in \eref{BBB}, solution \eref{z-sol} gives $z_i={ Q_{ij} x_j\over x_3}$. 
	Expanding the \eref{bubb}, we got the IBP recurrence relation
	\bea
	&&c_{m,n}i_{\lambda_0;m,n}+c_{m+1,n}i_{\lambda_0;m+1,n}+c_{m+1,n-1}i_{\lambda_0;m+1,n-1}+c_{m,n+1}i_{\lambda_0;m,n+1}\nn
	&&c_{m-1,n+1}i_{\lambda_0;m-1,n+1}+c_{m,n-1}i_{\lambda_0;m,n-1}+c_{m-1,n}i_{\lambda_0;m-1,n}+\delta_{2}=0~~~\label{ibpbubble}
	\eea
	where the $\delta_2$ is the boundary term, which we will compute later. Other  coefficients are
	\bea
	c_{m,n}&=&Q_{11}(1+m)+Q_{22}(1+n)+Q_{33}(1+\lambda_3)+\lambda_0\nn
	c_{m+1,n}&=&Q_{31}\lambda_3=-\lambda_3(a_2A_{11}+a_3A_{21}),~~~~c_{m+1,n-1}=Q_{21}n=-n(a_1A_{11}-a_3A_{31})\nn
	c_{m,n+1}&=&Q_{32}\lambda_3=-\lambda_3(a_2A_{12}-a_{3}A_{22}),~~~~c_{m-1,n+1}=Q_{12}m=m(a_1A_{22}+a_{2}A_{32})\nn
	c_{m,n-1}&=&Q_{23}n=-n(a_1A_{13}-a_3A_{33}),~~~~~c_{m-1,n}=Q_{13}m=m(a_1A_{32}+a_2A_{33})~~~\label{3.2}
	\eea
	Since we want to get the reduction of  $I_2(1,2)$, starting from $m=n=0$, we want to eliminate terms with
	indices $(m+1,n)$ and $(m+1,n-1)$, while keeping the term with index $(m,n+1)$. Thus we impose $c_{m+1,n}=0$ and  $c_{m+1,n-1}=0$,
	which can be satisfied by choosing the free parameters\footnote{For this example, one can check that we can not add another constraint to fix $a_1$.}
	\bea
	a_2&=&-\frac{a_1A_{21}}{A_{31}}=-a_1(m_1^2+m_2^2-p_1^2),~~~~~
	a_3=\frac{a_1A_{11}}{A_{31}}=2a_1m_1^2~~~\label{bub-a2a3}
	\eea
	After this choice, the matrix $\hat Q$ becomes to
	\bea
	\hat Q_{;r}=\left[\begin{array}{ccc}
	             \frac{1}{2}&\frac{a_1}{A_{31}}(A_{22}A_{31}-A_{21}A_{32})&\frac{a_1}{A_{31}}(A_{23}A_{31}-A_{21}A_{33})\nn
                 0&\frac{1}{2}-\frac{a_{1}}{A_{31}}(A_{12}A_{31}-A_{11}A_{32})&\frac{a_1}{A_{31}}(A_{11}A_{33}-A_{13}A_{31})\nn
                 0&\frac{a_1}{A_{31}}(A_{12}A_{21}-A_{11}A_{22})&\frac{1}{2}+\frac{a_1}{A_{31}}(A_{13}A_{21}-A_{11}A_{23})\end{array}\right]
	\eea
	 and it left us five terms with non-zero coefficients\footnote{where we use the convention $|\tilde A_{ij}|$ means the cofactor of matrix element $A_{ij}$}.
	\bea
	c_{m,n+1}&=&\frac{-a_1\lambda_3}{A_{31}}(A_{11}A_{22}-A_{12}A_{21})=\frac{-a_1\lambda_3}{A_{31}} |\tilde A_{33}|=a_1\lambda_3((m_1^2+m_2^2-p_1^2)^2-4m_1^2m_2^2)\nn
	c_{m-1,n+1}&=&-\frac{ma_1}{A_{31}}(A_{21}A_{32}-A_{22}A_{31})=-\frac{ma_1}{A_{31}}|\tilde A_{13}|=-a_1m(m_1^2-m_2^2-p_1^2)\nn
	c_{m,n-1}&=&\frac{na_1}{A_{31}}(A_{11}A_{33}-A_{13}A_{31})=\frac{na_1}{A_{31}}|\tilde A_{22}|=-a_1n\nn
	c_{m-1,n}&=&-\frac{ma_1}{A_{31}}(A_{21}A_{33}-A_{23}A_{31})=\frac{-ma_1}{A_{31}}|\tilde A_{12}|=a_1m\nn
	c_{m,n}&=&\frac{a_1}{A_{31}}\Big((1+n)(A_{11}A_{32}-A_{12}A_{31})-(\lambda_3+1)(A_{11}A_{23}-A_{13}A_{21})\Big)=\frac{a_1}{A_{31}}(n-\lambda_3)|\tilde A_{23}|\nn
	&=&\frac{a_1}{A_{31}}\Big((n-\lambda_3)(m_1^2-m_2^2+p_1^2)\Big)~~~\label{3.7}
	\eea

	{\bf The boundary $\delta_{2}$ term:}
	The $\delta_2$ term is given by
	\bea
	\delta_2&=&\sum_{i=1}^{3}\delta_{\lambda_i,0}\int d\Pi^{(2)} \Big\{z_iF^{\lambda_0}x_1^mx_2^nx_3^{\lambda_3+1}\Big\}|_{x_i=0}
	~~~~\label{Wang-bub-4}%
	\eea
	where the $\lambda_i$ represents the power of $x_i$. It is worth to emphasize that since $z_i$ contains $x_i$, 
	the total power $\lambda_i$ of $x_i$ is not equal to $m,n,\lambda_3$ in general.  
	Expanding it, we get\footnote{Since we have kept dimensional regularization $\eps$, the $\lambda_3$ can not be zero, thus the corresponding boundary term does not exist.}
	\bea
	\delta_2&=&\delta_{\lambda_1,0}\int d\Pi^{(2)}\Big(Q_{11}F^{\lambda_0}x_1^{m+1}x_2^nx_3^{\lambda_3}+Q_{12}F^{\lambda_0}x_1^mx_2^{n+1}x_3^{\lambda_3}+Q_{13}F^{\lambda_0}x_1^mx_2^nx_3^{\lambda_3+1}\Big)|_{x_1=0}\nn
	&&+\delta_{\lambda_2,0}\int d\Pi^{(2)}\Big(Q_{21}F^{\lambda_0}x_1^{m+1}x_2^nx_3^{\lambda_3}+Q_{22}F^{\lambda_0}x_1^{m}x_2^{n+1}x_3^{\lambda_3}+Q_{23}F^{\lambda_0}x_1^mx_2^nx_3^{\lambda_3+1}\Big)|_{x_2=0}~~~~\label{Wang-bub-5}
	\eea
	Remembering our extended notation explained under \eref{Wang-bub-2-6}, we have 
	\bea
	\int d\Pi^{(2)} F|_{x_1=0}^{\lambda_0}x_2^n\equiv&& i_{\lambda_0;-1,n},~~~~
	\int d\Pi^{(2)}F|_{x_2=0}^{\lambda_0}x_1^m\equiv  i_{\lambda_0;m,-1}~~~~\label{Wang-bub-6}
	\eea
	and the  $\delta_{2}$ term could be written as
	\bea
	\delta_{2;r}&=&\delta_{\lambda_1,0}\Big(Q_{11;r}i_{\lambda_0;m+1,n}+Q_{12;r}i_{\lambda_0;m,n+1}+Q_{13;r}i_{\lambda_0;m,n}\Big)\nn
	&&+\delta_{\lambda_2,0}\Big(Q_{21;r}i_{\lambda_0;m+1,n}+Q_{22;r}i_{\lambda_0;m,n+1}+Q_{23;r}i_{\lambda_0;m,n}\Big)\nn
	&=&\delta_{m,-1}Q_{11;r}i_{\lambda_0;-1,n}+\delta_{m,0}Q_{12;r}i_{\lambda_0;,-1,n+1}+\delta_{m,0}Q_{13;r}i_{\lambda_0;-1,n}\nn
	&&+\delta_{n,0}Q_{21;r}i_{\lambda_0;m+1,-1}+\delta_{n,-1}Q_{22;r}i_{\lambda_0;m,-1}+\delta_{n,0}Q_{23;r}i_{\lambda_0;m,-1}
	~~~\label{delta2r}
	\eea
	where the subscript $r$ in $\delta_{2;r}$ and $Q_{ij;r}$ means that the  $a_2$ and $a_3$ should be replaced by \eref{bub-a2a3}.\\
	Since the $m$ and $n$ could not be $-1$, the first and fifth terms are actually zero.
	
	%
	%
	%
	%
	Now we could use \eref{ibpbubble} and \eref{delta2r} to get our result directly. Setting  $m=0$ and $n=0$,
	all other terms in \eref{ibpbubble} are equal to zero, and we are left with\footnote{When setting $m=n=0$, except the 
		boundary term $\delta_2$, among  other
		seven terms in \eref{ibpbubble}, the coefficients of the second and the third terms have been chosen to be zero.
		For the other five terms, one can show that $c_{m-1,n+1}, c_{m,n-1}, c_{m-1, n}$ are zero by using the last line of \eref{3.2}. 
		There is another technical point. When $m=n=0$, the seventh term will contain $i_{\lambda_0;-1,0}$, which looks like the one
		defined in \eref{Wang-bub-6}. But they are, in fact, different. The one appeared in \eref{ibpbubble} with the measure $d\Pi^{(3)}$
		while the one appeared in \eref{Wang-bub-6} with measure $d\Pi^{(2)}$.}
	\bea
	c_{0,0}i_{\lambda_0;0,0}+c_{0,1}i_{\lambda_0;0,1}+\delta_{2;00}&=&0~~~~\label{Wang-bub-7}
	\eea
	with the coefficients
	\bea
	c_{0,0}&=&-a_1(D-3)(m_1^2-m_2^2+p_1^2)\nn
	c_{0,1}&=&a_1(D-3)\Big(m_1^4+m_2^4p_1^4-2m_1^2p_1^2-2m_2^2p_1^2-2m_1^2m_2^2\Big)\nn
	\delta_{2;00}&=&Q_{12;r}i_{\lambda_0;-1,1}+Q_{13;r}i_{\lambda_0;-1,0}+Q_{21;r}i_{\lambda_0;1,-1}+Q_{23;r}i_{\lambda_0;0,-1}
	~~~~\label{Wang-bub-8}
	\eea
	where
	\bea
	Q_{21;r}&=&\frac{-a_1}{A_{31}}(A_{21}A_{32}-A_{22}A_{31})=\frac{-a_1}{A_{31}}|\tilde A_{13}|,~~
	Q_{23;r}=\frac{-a_1}{A_{31}}(A_{11}A_{33}-A_{13}A_{31})=\frac{-a_1}{A_{31}}|\tilde A_{22}|\nn
	Q_{12;r}&=&\frac{-a_1}{A_{31}}(A_{21}A_{32}-A_{22}A_{31})=\frac{-a_{1}}{A_{31}}|\tilde A_{13}|,~~
	Q_{13;r}=\frac{-a_1}{A_{31}}(A_{21}A_{33}-A_{23}A_{31})=\frac{-a_1}{A_{31}}|\tilde A_{12}|
	\eea
	From it we could directly write down the answer %
	\bea
	i_{\lambda_0;0,1}&=&-\frac{c_{0,0}}{c_{0,1}}i_{\lambda_0;0,0}-\frac{Q_{21;r}}{c_{0,1}}i_{\lambda_0;1,-1}-\frac{Q_{23;r}}{c_{0,1}}i_{\lambda_0;0,-1}-\frac{Q_{12;r}}{c_{0,1}}i_{\lambda_0;-1,1}-\frac{Q_{13;r}}{c_{0,1}}i_{\lambda_0;-1,0}
	~~~~\label{Wang-bub-9}%
	\eea
	Translating back to scalar integrals, it is 
	\bea
	I_2(1,2)&=&c_{12\to11}I_2(1,1)+c_{12\to10}I_2(1,0)+c_{12\to20}I_2(2,0)+c_{12\to01}I_2(0,1)+c_{12\to02}I_2(0,2)~~~~\label{Wang-bub-10}
	\eea
	with $c_{12\to20}=0$ and   
	\bea
	c_{12\to11}&=&{-(-3 + D) (m_1^2 - m_2^2 + p_1^2))\over(
		m_1^4 + (m_2^2 - p_1^2)^2 - 2 m_1^2 (m_2^2 + p_1^2)},~~~
	c_{12\to10}=\frac{D-2}{-2 m_1^2 \left(m_2^2+p_1^2\right)+m_1^4+\left(m_2^2-p_1^2\right)^2}\nn
	c_{12\to01}&=&\frac{2-D}{-2 m_1^2 \left(m_2^2+p_1^2\right)+m_1^4+\left(m_2^2-p_1^2\right)^2},~~~
	c_{12\to02}=\frac{-m_1^2+m_2^2+p_1^2}{-2 m_1^2 \left(m_2^2+p_1^2\right)+m_1^4+\left(m_2^2-p_1^2\right)^2}~~~~~~~\label{Wang-bub-11}
	\eea
	Using 
	$I_2(2,0)=\frac{D-2}{2m_1^2}I_2(1,0)$\footnote{The reduction of tadpole with higher power is simple. Noticing that $I_2(1,0)\propto (m_1^2)^{\frac{D-2}{2}}$ by dimensional analysis, one can take the derivative over $m_1^2$ to get the wanted reduction coefficients. } and $
	I_2(0,2)=\frac{D-2}{2m_2^2}I_2(0,1)$ we have our final result of reduction of $I_2(1,2)$,
	\bea
	I_2(1,2)&=&c_{2\to2}I_2(1,1)+c_{2\to1;\bar2}I_2(1,0)+c_{2\to1;\bar1}I_2(0,1)~~~~\label{Wang-bub-2-14-1}
	\eea
	with the coefficients
	\bea
	c_{2\to2}&=&-\frac{(D-3) \left(m_1^2-m_2^2+p_1^2\right)}{-2 m_1^2 \left(m_2^2+p_1^2\right)+m_1^4+\left(m_2^2-p_1^2\right)^2}\nn
	c_{2\to1;\bar2}&=&\frac{D-2}{-2 m_1^2 \left(m_2^2+p_1^2\right)+m_1^4+\left(m_2^2-p_1^2\right)^2}\nn
	c_{2\to1;\bar1}&=&-\frac{(D-2) \left(m_1^2+m_2^2-p_1^2\right)}{2 m_2^2 \left(-2 m_1^2 \left(m_2^2+p_1^2\right)+m_1^4+\left(m_2^2-p_1^2\right)^2\right)}~~~~\label{Wang-bub-12}
	\eea
	which is given in \eref{Wang-bub-2-14}.
	\subsection{The general case of bubbles}
	Now let us consider the more complicated examples, i.e., the bubble with general higher power of propagators. By the choice \eref{bub-a2a3}
     we got an IBP recurrence realtion  \eref{3.7} and use it we could reduce the bubbles $i_{\lambda_0,m,n+1}$ to the simpler bubbles having  less total power of propagators and no higher power in $D_2$. Similarly, by choosing the different values of $a_2$ and $a_3$, we could get another IBP recurrence realtion to reduce the integral to those having no higher power in $D_1$. The choice is 
	%
	\bea
	a_2&=&-\frac{a_1A_{22}}{A_{32}},~~
	a_3=\frac{a_1A_{12}}{A_{32}}\nn
	\eea
	and the corresponding IBP recurrence is 
	\bea
	c_{m+1,n}i_{\lambda_0,m+1,n}+c_{m+1,n-1}i_{\lambda_0,m+1,n-1}+c_{m,n-1}i_{\lambda_0,m,n-1}+c_{m-1,n}i_{\lambda_0,m-1,n}+c_{m,n}i_{\lambda_0,m,n}+\delta_{2;r}&=&0\nn~~~~~\label{re2}
	\eea
	with the coefficients
	\bea
	c_{m+1,n}&=&(|\tilde A_{33}|)(D-3-m-n),~~c_{m+1,n-1}=-n|\tilde A_{23}|\nn
	c_{m,n-1}&=&n|\tilde A_{21}|,~~c_{m-1,n}=-m|\tilde A_{11}|,~~c_{m,n}=|\tilde A_{13}|(3+2m+n-D)
	\eea
	and  the boundary term
	\bea
	\delta_{2;r'}&=&-\delta_{m,0}|\tilde A_{11}|i_{\lambda_0,m,n}+\delta_{n,0}\Big(-|\tilde A_{32}|i_{\lambda_0,m+1,n}+|\tilde A_{21}|i_{\lambda_0,m,n}\Big)
	\eea
	Combining \eref{3.7} and \eref{re2}, we could reduce the general bubbles. 
	
	\subsubsection{The example: $I_2(1,3)$}
	In the example $I_2(1,3)$, we just need to reduce $D_2$ from power $3$ to $1$. The strategy is to use \eref{3.7} two times. 
	In the first step, by setting $m=0$ and $n=1$ in \eref{3.7} we got 
	\bea
	I_2(1,3)&=&\frac{|\tilde A_{23}|(D-5)}{2|\tilde A_{33}|}I_2(1,2)+\frac{|\tilde A_{22}|(D-3)}{2|\tilde A_{33}|}I_2(1,1)+\frac{-|\tilde A_{12}|(D-3)}{2|\tilde A_{33}|}I_2(0,2)+\frac{|\tilde A_{13}|}{|\tilde A_{33}|}I_2(0,3)~~~~~~\label{my13}
	\eea
	For the first term in \eref{my13}, setting $m=0$ and $n=0$ in \eref{3.7} again we have 
	\bea
	I_2(1,2)&=&\frac{|\tilde A_{23}|(D-3)}{|\tilde A_{33}|}I_2(1,1)+\frac{|\tilde A_{22}|(D-2)}{|\tilde A_{33}|}I_2(1,0)+\frac{|\tilde A_{13}|}{|\tilde A_{33}|}I_2(0,2)+\frac{-|\tilde A_{12}|(D-2)}{|\tilde A_{33}|}I_2(0,1)~~~~~~\label{my12}
	\eea
	Putting   \eref{my12} into \eref{my13} and using the reduction of tadpole\footnote{In general, we could repeat the similar procedure to give the tadpoles' IBP recurrence relation, and calculate them step by step. Here, for simplicity,  we could just use the trick, $I_2(1,0)\propto (m_1^2)^{\frac{D-2}{2}}$, and $I_2(0,1)\propto (m_2^2)^{\frac{D-2}{2}}$, to directly calculate the $I_2(2,0)=\frac{\d}{\d m_1^2}I_2(1,0)=\frac{D-2}{2m_1^2}I_2(1,0)$, $I_2(0,2)=\frac{\d}{\d m_2^2}I_2(0,1)=\frac{D-2}{2m_2^2}I_2(0,1)$, and $I_2(3,0)=\frac{1}{2}(\frac{\d}{\d m_1^2})^2 I_2(1,0)=\frac{(D-2)(D-4)}{8m_1^4}I_2(1,0)$, $I_2(0,3)=\frac{1}{2}(\frac{\d}{\d m_2^2})^2 I_2(0,3)=\frac{(D-2)(D-4)}{8m_2^4}I_2(0,1)$.}
	we get 
	\bea
	I_2(1,3)&=&c_{13\to11}I_2(1,1)+c_{13\to10}I_2(1,0)+c_{13\to01}I_2(0,1)
	\eea
	with the coefficients
	\bea
	c_{13\to11}&=&\frac{(|\tilde A_{23}||\tilde A_{33}|+|\tilde A_{23}|^2(D-5))(D-3)}{2|\tilde A_{33}|^2}\nn
	c_{13\to10}&=&\frac{|\tilde A_{22}||\tilde A_{23}|(D-5)(D-2)}{2|\tilde A_{33}|^2}\nn
	c_{13\to01}&=&\frac{(D-2)}{8|\tilde A_{33}|^2m_2^4}A_{21} (2 A_{32} |\tilde A_{23}| (D-5) m_{2}^2+A_{32} A_{t33} (D-4)-4 A_{33} |\tilde A_{23}| (D-5) m_{2}^4-2 A_{33}|\tilde A_{33}| (D-3) m_{2}^2)\nn
	&&-A_{22} A_{31} (2 |\tilde A_{23}| (D-5) m_{2}^2+A_{t33} (D-4))+2 A_{23} A_{31} m_{2}^2 (2 |\tilde A_{23}| (D-5) m_{2}^2+|\tilde A_{33}| (D-3))
	\eea
	The result is confirmed with  FIRE6. In this example, we just need to solve 2 equations in reducing bubbles' topology. 
	%
	\subsubsection{The example: $I_2(3,5)$}
	For this example we need to  use \eref{re2} to lower the power of $D_1$ and  \eref{3.7} to lower the power of $D_2$. 
	Setting  $m=1$ and $n=4$ in \eref{re2} we can reduce $I_2(3,5)$ to $I_2(2,4)$, $I_2(2,5)$, $I_2(1,5)$ and $I_2(3,4)$.
	\bea
	I_2(3,5)&=&\frac{|\tilde A_{11}|(D-7)}{2|\tilde A_{33}|}I_2(1,5)+\frac{-|\tilde A_{13}|(D-9)}{2|\tilde A_{33}|}I_2(2,5)+\frac{-|\tilde A_{21}|(D-7)}{2|\tilde A_{33}|}I_2(2,4)+\frac{|\tilde A_{23}|}{|\tilde A_{33}|}I_2(3,4)~~~~~~
	\eea
	Then setting $m=1$ and $n=3$ in \eref{re2}, we reduce $I_2(3,4)$ to $I_2(1,4)$, $I_2(2,3)$, $I_2(2,4)$ and $I_2(3,3)$.
	\bea
	I_2(3,4)&=&\frac{-|\tilde A_{23}|}{|\tilde A_{33}|}I_2(3,3)+\frac{-|\tilde A_{13}|(D-8)}{2|\tilde A_{33}|}I_2(2,4)+\frac{-|\tilde A_{21}|(D-6)}{2|\tilde A_{33}|}I_2(2,3)+\frac{|\tilde A_{11}|(D-6)}{2|\tilde A_{33}|}I_2(1,4)~~~~~~~~~
	\eea
	With the same idea going down, we just need to solve $14$ equation to complete reduce the $I_{2}(3,5)$. The analytic expression by these $14$ equations
	have also been confirmed by FIRE6. 
	%
	\subsection{The triangle's case}
	%
	The triangle $I_3(m+1,n+1,q+1)$ is given by
	\bea
	I_3(m+1,n+1,q+1)&=&\int \frac{d^{D}l}{(l^2-m_1^2)^{m+1}((l-p_1)^2-m_2^2)^{n+1}((l+p_3)^2-m_3^2)^{q+1}}~~~\label{Wang-tri-1}
	\eea
	The parametric form of it is
	\bea
	I_3(m+1,n+1,q+1)&=&i(-1)^{3+m+n+q}\frac{\Gamma(-\lambda_0)}{\Gamma(m+1)\Gamma(n+1)\Gamma(q+1)\Gamma(\lambda_4+1)}i_{\lambda_0,m,n,q}
	~~~\label{Wang-tri-2}
	\eea
	where
	\bea
	i_{\lambda_0;m,n,q}&=&\int d\Pi^{(4)}F^{\lambda_0}x_1^mx_2^nx_3^qx_4^{\lambda_4},~~\lambda_0=-\frac{D}{2},~~\lambda_4=-4-2\lambda_0-m-n-q=D-4-m-n-q
	~~~~~~\label{Wang-tri-3}
	\eea
	Using the expression \eref{Wang-2-8}, we have
	\bea
	U(x)&=&x_1+x_2+x_3,~~~~~
	V(x)=x_1x_2 p_1^2+x_1x_3 p_3^2+x_2x_3 p_2^2\nn
	f(x)&=&-V+U\sum m_i^2x_i=(x_1+x_2+x_3)(x_1m_1^2+x_2m_2^2+x_3m_3^2)-x_1x_2p_1^2-x_2x_3p_2^2-x_1x_3p_3^2\nn
	F(x)&=&U(x)x_4+f(x)\nn
	&=&\Big(x_1+x_2+x_3\Big)\Big(m_1^2x_1+m_2^2x_2+m_3^2x_3+x_4\Big)-x_1x_2p_1^2-x_2x_3p_2^2-x_1x_3p_3^2
	~~~~\label{fi-xi}
	\eea
	Thus we can read out matrices 
	\bea
	\hat A&=&\left[\begin{array}{cccc}
		2m_1^2&m_1^2+m_2^2-p_1^2&m_1^2+m_3^2-p_3^2&1\nn
		m_1^2+m_2^2-p_1^2&2m_2^2&m_2^2+m_3^2-p_2^2&1\nn
		m_1^2+m_3^2-p_3^2&m_2^2+m_3^2-p_2^2&2m_3^2&1\nn
		1&1&1&0
	\end{array}\right]
	~~~\hat K_{A}=
	\left[\begin{array}{cccc}
		0&a_1&a_2&a_3\nn
		-a_1&0&a_4&a_5\nn
		-a_2&-a_4&0&a_6\nn
		-a_3&-a_5&-a_6&0
	\end{array}\right]
	\nn
	\hat Q&=&\frac{1}{2}\hat I+\hat K_{A}\hat A~~~~~~\label{Wang-tri-4}
	\eea
	%
	
	\subsubsection{Deriving the recurrence relation}
	Taking $B=\frac{-1}{x_4}$ in \eref{z-sol}, we got $z_i=\frac{Q_{ij} x_j}{x_4}$.
	Taking this relation into our IBP identities \eref{bubb}, we got
	\bea
	\sum_{i=1}^4 \int d\Pi^{(4)} \Big\{z_iF^{\lambda_0}x_1^mx_2^nx_3^qx_4^{\lambda_4+1}\Big\}+\delta_{3}&=&0~~~~~~\label{Wang-tri-5}
	\eea
	where we will deal with the boundary $\delta_3$ term later. After expanding the first term, we got
	we got
	\bea
	&&c_{m,n,q}i_{\lambda_0;m,n,q}+c_{m+1,n,q}i_{\lambda_0;m+1,n,q}+c_{m+1,n,q-1}i_{\lambda_0;m+1,n,q-1}+c_{m+1,n-1,q}i_{\lambda_0;m+1,n-1,q}\nn
	&&+c_{m-1,n+1,q}i_{\lambda_0;m-1,q+1,q}+c_{m,n+1,q-1}i_{\lambda_0;m,n+1,q-1}+c_{m,n+1,q}i_{\lambda_0;m,n+1,q}+c_{m,n,q+1}i_{\lambda_0;m,n,q+1}\nn
	&&+c_{m,n-1,q+1}i_{\lambda_0;m,n-1,q+1}+c_{m-1,n,q+1}i_{\lambda_0;m-1,n,q+1}+c_{m-1,n,q}i_{\lambda_0;m-1,n,q}+c_{m,n-1,q}i_{m,n-1,q}\nn
	&&+c_{m,n,q-1}i_{\lambda_0;m,n,q-1}+\delta_3=0~~~~~~\label{Wang-tri-6}
	\eea
	with the coefficients
	\bea
	c_{m,n,q}&=&\lambda_0+(m+1) Q_{11}+(n+1) Q_{22}+(q+1) Q_{33}+(\lambda_4+1) Q_{44}\nn
	c_{m+1,n,q}&=&\lambda_4 Q_{41},~~c_{m+1,n,q-1}=q Q_{31},~~c_{m+1,n-1,q}=n Q_{21},~~c_{m,n,q-1}=q Q_{34}\nn
	c_{m-1,n+1,q}&=&m Q_{12},~~c_{m,n+1,q-1}=q Q_{32},~~c_{m,n+1,q}=\lambda_4 Q_{42},~~c_{m,n,q+1}=\lambda_4 Q_{43}\nn
	c_{m,n-1,q+1}&=&n Q_{23},~~c_{m-1,n,q+1}=m Q_{13},~~c_{m-1,n,q}=m Q_{14},~~c_{m,n-1,q}=n Q_{24}~~~~~~\label{Wang-tri-7}
	\eea
	Now, we could choose particular value of our six parameters, $a_1$, $a_2$, $a_3$, $a_4$, $a_5$, $a_6$ to let the coefficients $c_{m+1,n,q}$, $c_{m+1,n,q-1}$, $c_{m+1,n-1,q}$, $c_{m-1,n+1,q}$, $c_{m,n+1,q}$, $c_{m,n+1,q}$ be zero. The solution is
	\bea
	a_2&=&-a_1\frac{A_{21}A_{42}-A_{22}A_{41}}{A_{31}A_{42}-A_{32}A_{41}}=-\frac{a_1 \left(-m_1^2+m_2^2+p_1^2\right)}{-m_1^2+m_2^2+2 (p_1\cdot p_2)+p_1^2}\nn
	a_3&=&\frac{a_{1} (A_{21} A_{32}-A_{22} A_{31})}{A_{31} A_{42}-A_{32} A_{41}}=-\frac{a_1 \left(m_1^2-m_2^2-p_1^2\right) \left(m_2^2+m_3^2-p_2^2\right)}{-m_1^2+m_2^2+2 (p_1\cdot p_2)+p_1^2}-2 a_1 m_2^2\nn
	a_4&=&\frac{a_{1} (A_{11} A_{42}-A_{12} A_{41})}{A_{31} A_{42}-A_{32} A_{41}}=-\frac{a_1 \left(m_1^2-m_2^2+p_1^2\right)}{-m_1^2+m_2^2+2 (p_1\cdot p_2)+p_1^2}\nn
	a_5&=&\frac{-a_{1} (A_{11} A_{32}-A_{12} A_{31})}{A_{31} A_{42}-A_{32} A_{41}}=\frac{a_1 \left(m_1^2-m_2^2+p_1^2\right) \left(m_1^2+m_3^2-2 (p_1\cdot p_2)-p_1^2-p_2^2\right)}{-m_1^2+m_2^2+2 (p_1\cdot p_2)+p_1^2}+2 a_1 m_1^2\nn
	a_6&=&\frac{a_{1} (A_{11} A_{22}-A_{12} A_{21})}{A_{31} A_{42}-A_{32} A_{41}}=\frac{a_1 \left(m_1^4-2 m_1^2 \left(m_2^2+p_1^2\right)+\left(m_2^2-p_1^2\right)^2\right)}{-m_1^2+m_2^2+2 (p_1\cdot p_2)+p_1^2}~~~\label{a2toa6}
	\eea
	Then the matrix $\hat Q$ becomes to
	\bea
	\hat Q_{r}&=&\frac{1}{\Delta_{A}}\left[\begin{array}{cccc}
		\frac{1}{2}\Delta_{A}&0&a_1|\tilde A_{14}|&~~~~~~~a_1|\tilde A_{13}|\nn
		0&\frac{1}{2}\Delta_{A}&-a_1|\tilde A_{24}|&~~~~~~~a_1|\tilde A_{23}|\nn
		0&0&\frac{1}{2}\Delta_{A}+a_1|\tilde A_{34}|&~~~~~~~a_1|\tilde A_{33}|\nn
		0&0&-a_1|\tilde A_{44}|&~~~~\frac{1}{2}\Delta_A -a_1|\tilde A_{43}|
	\end{array}\right],~~~\Delta_{A}=Det\left[\begin{array}{cc}
		A_{31}&A_{32}\nn
		A_{41}&A_{42}   \end{array}\right]=A_{31}A_{42}-A_{32}A_{41}
	\eea
	After this, we have the reduced  IBP relation with only the propagator $D_3=(l+p_3)^2-m_3^2$ having one increasing power
	\bea
	&&c_{m,n,q}i_{\lambda_0;m,n,q}+c_{m,n,q+1}i_{\lambda_0;m,n,q+1}+c_{m,n-1,q+1}i_{\lambda_0;m,n-1,q+1}+c_{m-1,n,q+1}i_{\lambda_0;m-1,n,q+1}\nn
	&&+c_{m-1,n,q}i_{\lambda_0;m-1,n,q}+c_{m,n-1,q}i_{\lambda_0;m,n-1,q}+c_{m,n,q-1}i_{\lambda_0;m,n,q-1}+\delta_{3;r}=0~~~~~\label{re-ibp}
	\eea
	with the coefficients
	\bea
	c_{m,n,q}&=&\lambda_0+mQ_{11;r}+nQ_{22;r}+qQ_{33;r}+Q_{11;r}+Q_{22;r}+Q_{33;r}+\lambda_4Q_{44;r}+Q_{44;r}\nn
	c_{m,n,q+1}&=&\lambda_4 Q_{43;r}=\frac{-a_1\lambda_4}{A_{31}A_{42}-A_{32}A_{41}}|\tilde A_{44}|,~~c_{m,n-1;q+1}=nQ_{23;r}=\frac{-a_1n}{A_{31}A_{42}-A_{32}A_{41}}|\tilde A_{24}|\nn
	c_{m-1,n,q+1}&=&mQ_{13;r}=\frac{a_1m}{A_{31}A_{42}-A_{32}A_{41}}|\tilde A_{14}|,~~c_{m-1,n,q}=mQ_{14;r}\frac{a_1m}{A_{31}A_{42}-A_{32}A_{41}}|\tilde A_{13}|\nn
	c_{m,n-1,q}&=&nQ_{24;r}=\frac{-a_1n}{A_{31}A_{42}-A_{32}A_{41}}|\tilde A_{23}|,~~c_{m,n,q-1}=qQ_{34;r}=\frac{a_1q}{A_{31}A_{42}-A_{23}A_{41}}|\tilde A_{33}|~~~~~~\label{Wang-tri-8}
	\eea
	where the subscript $r$ in $\delta_{3;r}$ and $Q_{ij;r}$ means that the  parameters $a_2$ to $a_6$  should be replaced by \eref{a2toa6}.

	{\bf The reduction of the boundary $\delta_{3}$ part:}
	Similarly to the bubble's situation, taking the value of $z_i$ into the $\delta_{3}$ part, we have the result 
	\bea
	\delta_{3;r}&=&\Big(\delta_{m+1,0}Q_{11;r}+\delta_{m,0}Q_{14;r}\Big)i_{\lambda_0,-1,n,q}+\delta_{m,0}Q_{12;r}i_{\lambda_0,-1,n+1,q}+\delta_{m,0}Q_{13;r}i_{\lambda_0,-1,n,q+1}\nn
	&&+\delta_{n,0}Q_{21;r} i_{\lambda_0,m+1,-1,q}+\Big(\delta_{n+1,0}Q_{22;r}+\delta_{n,0}Q_{24;r}\Big)i_{\lambda_0,m,-1,q}+\delta_{n,0}Q_{23;r} i_{\lambda_0,m,-1,q+1}\nn
	&&+\delta_{q,0}Q_{31;r}i_{\lambda_0,m+1,n,-1}+\delta_{q,0}Q_{32;r}i_{\lambda_0,m,n+1,-1}+\Big(\delta_{q+1,0}Q_{33;r}+\delta_{q,0}Q_{34;r}\Big)i_{\lambda_0,m,n,-1}
	~~~~~~\label{Wang-tri-9}%
	\eea
	the $i_{\lambda_0,m,n,-1}$, $i_{\lambda_0,m,-1,q}$ and $i_{\lambda_0,-1,n,q}$ contribute to the sub-topology of triangle, i.e. the bubble\footnote{Since the boundary term having only one $x_i=0$, it reduces to the sub-topologies with only one propagator pinched. }.
	%
	\subsubsection{The triangle's example: $I_3(1,1,2)$}
	%
	%
	Now we apply the complete recurrence relation to the  example $I_3(1,1,2)$. 
	Setting $m=n=q=0$ in \eref{re-ibp}, we got our recurrence relation.
	\bea
	c_{0,0,0}i_{\lambda_0,0,0,0}+c_{0,0,1}i_{\lambda_0,0,0,1}+\delta_{3;000}&=&0~~~\label{relationi3112}
	\eea
	with the coefficients
	%
	\bea
	c_{0,0,1}&=&\lambda_4 Q_{43;r}=-\frac{1}{-m_1^2+m_2^2+2 (p_1\cdot p_2)+p_1^2}\times\Big\{2 a_1 (D-4) \Big(m_1^4 p_2^2-2 m_1^2 \big(m_2^2 ((p_1\cdot p_2)+p_2^2)-m_3^2 (p_1\cdot p_2)\nn
	&&+p_2^2 ((p_1\cdot p_2)+p_1^2)\big)+m_2^4 (2 (p_1\cdot p_2)+p_1^2+p_2^2)+m_2^2 \big(2 (p_1\cdot p_2) (2 (p_1\cdot p_2)+p_1^2+p_2^2)\nn
	&&-2 m_3^2 ((p_1\cdot p_2)+p_1^2)\big)+p_1^2 (m_3^4-2 m_3^2 ((p_1\cdot p_2)+p_2^2)+p_2^2 (2 (p_1\cdot p_2)+p_1^2+p_2^2))\Big)\Big\}\nn
	c_{0,0,0}&=&-\frac{D}{2}+Q_{11;r}+Q_{22;r}+Q_{33;r}+(D-3)Q_{44;r}\nn
	&=&-\frac{2 a_1 (D-4) \left(m_1^2 (p_1\cdot p_2)-m_2^2 ((p_1\cdot p_2)+p_1^2)+p_1^2 \left(m_3^2-(p_1\cdot p_2)-p_2^2\right)\right)}{-m_1^2+m_2^2+2 (p_1\cdot p_2)+p_1^2}~~~~~~\label{Wang-tri-10}
	\eea
	One can see that in \eref{relationi3112}, only two terms of triangle topologies are left: one is the scalar basis and one is the 
	target we want to reduce. Other five terms in \eref{re-ibp} disappear by the expression in \eref{Wang-tri-8}.  Thus there is no need to solve mixed IBP relations.
	The $\delta_{3}$ term becomes to
	\bea
	\delta_{p;000}\equiv  \delta_{p;r}|_{m=0,n=0,q=0}&=&Q_{14;r}i_{\lambda_0,-1,0,0}+Q_{12;r}i_{\lambda_0,-1,1,0}+Q_{13;r}i_{\lambda_0,-1,0,1}\nn
	&&+Q_{21;r}i_{\lambda_0,1,-1,0}+Q_{24;r}i_{\lambda_0,0,-1,0}+Q_{23;r}i_{\lambda_0,0,-1,1}\nn
	&&+Q_{31;r}i_{\lambda_0,1,0,-1}+Q_{32;r}i_{\lambda_0,0,1,-1}+Q_{34;r}i_{\lambda_0,0,0,-1}~~~\label{dp000}
	\eea
	Translating back to the form of $I$, 
	We have the result
	\bea
	I_3(1,1,2)&=&c_{3\to111}I_3(1,1,1)+c_{3\to110}I_3(1,1,0)+c_{3\to101}I_3(1,0,1)+c_{3\to011}I_3(0,1,1)\nn
	&&c_{3\to210}I_3(2,1,0)+c_{3\to201}I_3(2,0,1)+c_{3\to120}I_3(1,2,0)+c_{3\to021}I_3(0,2,1)\nn
	&&c_{3\to102}I_{3}(1,0,2)+c_{3\to012}I_3(0,1,2)~~~~~~\label{Wang-tri-11}
	\eea
	with the coefficients
	\bea
	c_{3\to111}&=&\frac{c_{0,0,0}\Gamma(D-3)}{c_{0,0,1}\Gamma(D-4)},~~c_{3\to110}=-\frac{Q_{34;r} \Gamma (D-2)}{c_{0,0,1} \Gamma (D-4)},~~c_{3\to101}=-\frac{Q_{24;r} \Gamma (D-2)}{c_{0,0,1} \Gamma (D-4)},~~c_{3\to011}=-\frac{Q_{14;r} \Gamma (D-2)}{c_{0,0,1} \Gamma (D-4)}\nn
	c_{3\to210}&=&\frac{Q_{31;r} \Gamma (D-3)}{c_{0,0,1} \Gamma (D-4)},~~c_{3\to201}=\frac{Q_{21;r} \Gamma (D-3)}{c_{0,0,1} \Gamma (D-4)},~~c_{3\to021}=\frac{Q_{12;r} \Gamma (D-3)}{c_{0,0,1} \Gamma (D-4)},~~c_{3\to120}=\frac{Q_{32;r} \Gamma (D-3)}{c_{0,0,1} \Gamma (D-4)}\nn
	c_{3\to102}&=&\frac{Q_{23;r} \Gamma (D-3)}{c_{0,0,1} \Gamma (D-4)},~~c_{3\to012}=\frac{Q_{13;r} \Gamma (D-3)}{c_{0,0,1} \Gamma (D-4)}
	~~~~~~\label{Wang-tri-12}%
	\eea
	The last step is to reduce the bubbles with one propagator having power two. This problem has been solved in the
	previous subsection (see  \eref{Wang-bub-2-14-1}). 
	With proper relabeling of external variables for last six terms in \eref{Wang-tri-11} and collecting all coefficients together, we have 
	we got
	\bea
	I_3(1,1,2)&=&c_{3\to3}I_3(1,1,1)+c_{3\to2;\bar3}I_3(1,1,0)+c_{3\to2;\bar2}I_3(1,0,1)+c_{3\to2;\bar1}I_3(0,1,1)\nn
	&&+c_{3\to1;\bar2\bar3}I_3(1,0,0)+c_{3\to1;\bar1\bar3}I_3(0,1,0)+c_{3\to1;\bar1\bar2}I_3(0,0,1)~~~~~~\label{Wang-tri-13}
	\eea
	Since the explicit expressions of these coefficients are long, we have given them in the companion Mathematica notebook. 
	The result is  confirmed  by FIRE6.
	 \subsubsection{The general case in triangles}
	Similarly to the bubbles' case, by different choices, we could get three  IBP recurrence relations, where in each one  only one term has  one propagator having higher power. For simplicity, let us label the IBP recurrence relation $eq_i$ which shifting the propagator $D_i$. Now we could use the $eq_{i}$ with $i=1,2,3$ to calculate the general case of triangles. Let us denote
	\bea
	&&eq_1:\Big(a_{1^{+}}1^{+}+a_{1^+3^-}1^+3^-+a_{1^+2^-}1^+2^-+a_{3^-}3^-+a_{2^-}2^-+a_{1^-}1^-+a_{0}\Big)i_{\lambda_0,m,n,q}+\delta_{3;r,eq1}=0\nn
	&&eq_2:\Big(b_{2^+}2^++b_{2^+3^-}2^+3^-+b_{1^-2^+}1^-2^++b_{3^-}3^-+b_{2^-}2^-+b_{1^-}1^-+b_{0}\Big)i_{\lambda_0,m,n,q}+\delta_{3;r,eq2}=0\nn
	&&eq_3:\Big(c_{3^+}3^++c_{2^-3^+}2^-3^++c_{1^-3^+}1^-3^++c_{3^-}3^-+c_{2^-}2^-+c_{1^-}1^-+c_{0}\Big)i_{\lambda_0,m,n,q}+\delta_{3;r,eq3}=0~~~
	\eea
	with all  coefficients having the same form as \eref{re-ibp}.  Combining them all, we could reduce the general triangles.
	For example, for $I_3(2,2,3)$,  starting  with 
	seting $m=0$, $n=1$ and $q=2$ in $eq_1$,  we could reduce $I_3(2,2,3)$ to $I_3(1,1,3)$, $I_3(1,2,2)$, $I_3(1,2,3)$, $I_3(2,1,3)$ and $I_3(2,2,2)$ and boundary terms, the general bubbles. Then seting $m=0$, $n=0$ and $q=2$ in $eq_1$, we could reduce $I_3(2,1,3)$ to $I_3(1,1,2)$, $I_3(1,1,3)$, $I_3(2,1,2)$.   After {\bf 12} steps, we got the result of the reduction of the  triangle's topology. The boundary terms involve bubbles
	and tadpoles, which have been dealt in previous subsections. Finally, we could get all the coefficients from $I_3(2,2,3)$ to all the scalar basis.
	 
	%
	%
	\subsection{The box case }
	%
	The general form of  box is given by
	\bea
	I_4(n_1+1,n_2+1,n_3+1,n_4+1)&=&\int \frac{d^{D}l}{D_1^{n_1+1}D_2^{n_2+1}D_3^{n_3+1}D_4^{n_4+1}}~~~~\label{Wang-box-1}
	\eea
	with
	\bea
	D_1&=&l^2-m_1^2,~~
	D_2=(l-p_1)^2-m_2^2,~~
	D_3=(l-p_1-p_2)^2-m_3^2,~~
	D_4=(l+p_4)^2-m_4^2~~~~\label{Wang-box-2}
	\eea
	The parametric form of $I_4(n_1+1,n_2+1,n_3+1,n_4+1)$ could be written as
	\bea
	& & I_4(n_1+1,n_2+1,n_3+1,n_4+1)=\frac{i(-1)^{4+n_1+n_2+n_3+n_4}\Gamma(-\lambda_0)}{\Gamma(n_1+1)\Gamma(n_2+1)\Gamma(n_3+1)\Gamma(n_4+1)\Gamma(\lambda_5+1)}i_{\lambda_0;n_1,n_2,n_3,n_4}~~~~\label{Wang-box-3}
	\eea
	where
	\bea
	i_{\lambda_0;n_1,n_2,n_3,n_4}&=&\int d\Pi^{(5)} F^{\lambda_0}x_1^{n_1}x_2^{n_2}x_3^{n_3}x_4^{n_4}x_5^{\lambda_5}
	=\int d\Pi^{(5)} (Ux_5+f)^{\lambda_0}x_1^{n_1}x_2^{n_2}x_3^{n_3}x_4^{n_4}x_5^{\lambda_5}\nn
	d\Pi^{(5)}&=&dx_1dx_2dx_3dx_4dx_5\delta(\sum x_j-1),~~~~~~~
	\lambda_0=-\frac{D}{2}\nn
	\lambda_5&=&-5-n_1-n_2-n_3-n_4-2\lambda_0=D-5-n_1-n_2-n_3-n_4~~~~\label{Wang-box-4}
	\eea
	and the functions are 
	\bea
	U(x)&=&x_1+x_2+x_3+x_4\nn
	V(x)&=&x_1x_2p_1^2+x_1x_3(p_1+p_2)^2+x_1x_4(p_1+p_2+p_3)^2+x_2x_3p_2^2+x_2x_4(p_2+p_3)^2+x_3x_4p_3^2\nn
	f(x)&=&-V(x)+U(x)\sum m_i^2 x_i\nn
	&=&m_1^2x_1+m_2^2x_2+m_3^2x_3+m_4^2x_4+(m_1^2+m_2^2-p_1^2)x_1x_2\nn
	&&+[m_1^2+m_3^2-(p_1+p_2)^2]x_1x_3+[m_1^2+m_4^2-(p_1+p_2+p_3)^2]x_1x_4\nn
	&&+(m_2^2+m_3^2-p_2^2)x_2x_3+[m_2^2+m_4^2-(p_2+p_3)^2]x_2x_4+(m_3^2+m_4^2-p_3^2)x_3x_4\nn
	F(x)&=&U(x) x_5+f(x)\nn
	&=&m_1^2x_1^2+m_2^2x_2^2+m_3^2x_3^2+m_4^2x_4^2\nn
	&&+(m_1^2+m_2^2-p_1^2)x_1x_2+[m_1^2+m_3^2-(p_1+p_2)^2]x_1x_3+[m_1^2+m_4^2-(p_1+p_2+p_3)^2]x_1x_4\nn
	&&+[m_2^2+m_3^2-p_2^2]x_2x_3+[m_2^2+m_4^2-(p_2+p_3)^2]x_2x_4+[m_3^2+m_4^2-p_3^2]x_3x_4\nn
	&&+x_1x_5+x_2x_5+x_3x_5+x_4x_5\nn
	&=&(x_1+x_2+x_3+x_4)(m_1^2x_1+m_2^2x_2+m_3^2x_3+m_4^2x_4+x_5)\nn
	&&-x_1x_2p_1^2-x_1x_3(p_1+p_2)^2-x_1x_4(p_1+p_2+p_3)^2-x_2x_3p_2^2-x_2x_4(p_2+p_3)^2-x_3x_4p_3^2~~~~\label{Wang-box-5}
	\eea
	Now the matrix  are given by
	\bea
	\hat A&=&\left[\begin{array}{ccccc}
		2m_1^2~&m_1^2+m_2^2-p_1^2~&m_1^2+m_3^2-p_{12}^2~&m_1^2+m_4^2-p_{13}^2~&1\nn
		m_1^2+m_2^2-p_1^2~&2m_2^2~&m_2^2+m_3^2-p_2^2~&m_2^2+m_4^2-p_{23}^2~&1\nn
		m_1^2+m_3^2-p_{12}^2~&m_2^2+m_3^2-p_2^2~&2m_3^2~&m_3^2+m_4^2-p_3^2~&1\nn
		m_1^2+m_4^2-p_{13}^2~&m_2^2+m_4^2-p_{23}^2~&m_3^2+m_4^2-p_3^2~&2m_4^2~&1\nn
		1~~&1~~&1~~&1~~&0
	\end{array}\right],~~
	K_A=\left[\begin{array}{ccccc}
		0&a_1&a_2&a_3&a_4\nn
		-a_1&0&a_5&a_6&a_7\nn
		-a_2&-a_5&0&a_8&a_9\nn
		-a_3&-a_6&-a_8&0&a_{10}\nn
		-a_4&-a_7&-a_9&-a_{10}&0
	\end{array}\right]\nn~~~~\label{Wang-box-6}
	\eea
	where $p_{ij}\equiv p_i+p_{i+1}\cdots p_j$.
	%
	%
	%
	%
	%
	\subsubsection{Deriving the recurrence relation}
	Taking $B=\frac{-1}{x_5}$ in \eref{z-sol} we got
	\bea
	&&\Big\{c_{n_1+1,n_2,n_2,n_4}1^{+}+c_{n_1+1,n_2,n_3,n_4-1}`^{+}4^{-}+c_{n_1+1,n_2,n_3-1,n_4}1^{+}3^{-}+c_{n_1+1,n_2-1,n_3,n_4}1^{+}2^{-}\nn
	&&+c_{n_1,n_2+1,n_3,n_4}2^{+}+c_{n_1,n_2+1,n_3,n_4-1}2^{+}4^{-}+c_{n_1,n_2+1,n_3-1,n_4}2^{+}3^{-}+c_{n_1-1,n_2+1,n_3,n_4}2^{+}1^{-}\nn
	&&+c_{n_1,n_2,n_3+1,n_4}3^{+}+c_{n_1,n_2,n_3+1,n_4-1}3^{+}4^{-}+c_{n_1,n_2-1,n_3+1,n_4}3^{+}2^{-}+c_{n_1-1,n_2,n_3+1,n_4}3^{+}1^{-}\nn
	&&+c_{n_1,n_2,n_3,n_4+1}4^{+}+c_{n_1,n_2,n_3-1,n_4+1}4^{+}3^{-}+c_{n_1,n_2-1,n_3,n_4+1}4^{+}2^{-}+c_{n_1-1,n_2,n_3,n_4+1}4^{+}1^{-}\nn
	&&+c_{n_1,n_2,n_3,n_4-1}4^{-}+c_{n_1,n_2,n_3-1,n_4}3^{-}+c_{n_1,n_2-1,n_3,n_4}2^{-}+c_{n_1-1,n_2,n_3,n_4}1^{-}+c_{n_1,n_2,n_3,n_4}
	\Big\}i_{n_1,n_2,n_3,n_4}+\delta_{4}=0~~~\label{box111}\nn
	\eea
	where 
	\bea
	j^{+}i_{n_1\cdots n_j\cdots n_k}&=&i_{n_1\cdots n_j+1\cdots n_k},~~~~~j^{-}i_{n_1\cdots n_j\cdots n_k}
	=i_{n_1\cdots n_j-1\cdots n_k}~~~~\label{Wang-box-7}
	\eea
	Similarly, we could choose particular value of the parameters $a_2$ to $a_{10}$ with $a_1$ free to make the coefficients of  terms in the first three lines of \eref{box111} be zero.  The analytic solution will be collected in the companion Mathematica notebook, while we could express the solution of the parameters by matrix elements of $\hat A$. 
	\bea
	a_2&=&\frac{-a_1}{\Delta_{box}}|\tilde A_{13,45}|,~~~
	a_{3}=\frac{a_1}{\Delta_{Box}}|\tilde A_{14,45}|,~~~
	a_4=\frac{-a_1}{\Delta_{Box}}|\tilde A_{15,45}|,~~~
	a_5=\frac{a_1}{\Delta_{Box}}|\tilde A_{23,45}|,~~~
	a_6=\frac{-a_1}{\Delta_{Box}}|\tilde A_{24,45}|\nn
	a_{7}&=&\frac{a_1}{\Delta_{box}}|\tilde A_{25,45}|,~~~a_{8}=\frac{a_1}{\Delta_{box}}|\tilde A_{34,45}|,~~~a_9=\frac{-a_1}{\Delta_{Box}}|\tilde A_{35,45}|,~~~a_{10}=\frac{a_1}{\Delta_{box}}|\tilde A_{45,45}|,
	\Delta_{Box}=\left|\begin{array}{ccc}
		A_{31}&A_{32}&A_{33}\nn
		A_{41}&A_{42}&A_{43}\nn
		A_{51}&A_{52}&A_{53}
	\end{array}\right|
	\eea
	where $|\tilde A_{ij,kl}|$ means the determinant of the matrix $A$ after we removed the $i,j$th rows and $k,l$th columns.Then the matrix $\hat Q$ becomes to
	\bea
	\hat Q_{r}&=&\frac{1}{\Delta_{Box}}\left[\begin{array}{ccccc}
		\frac{1}{2}\Delta_{Box}&0&0&-a_1|\tilde A_{15}|&-a_1|\tilde A_{14}|\nn
		0&\frac{1}{2}\Delta_{Box}&0&a_1|\tilde A_{25}|&a_1|\tilde A_{24}|\nn
		0&0&\frac{1}{2}\Delta_{Box}&-a_{1}|\tilde A_{35}|&-a_{1}|\tilde A_{34}|\nn
		0&0&0&\frac{1}{2}\Delta_{Box}+a_{1}|\tilde A_{45}|&a_1|\tilde A_{44}|\nn
		0&0&0&-a_1|\tilde A_{55}|&\frac{1}{2}\Delta_{Box}-a_1|\tilde A_{54}|
	\end{array}\right]
	\eea
	After this we got the simplified recurrence relation
	\bea
	&&c_{n_1,n_2,n_3,n_4+1}i_{n_1,n_2,n_3,n_4+1}+c_{n_1,n_2,n_3-1,n_4+1}i_{n_1,n_2,n_3-1,n_4+1}\nn
	&&+c_{n_1,n_2-1,n_3,n_4+1}i_{n_1,n_2-1,n_3,n_4+1}+c_{n_1-1,n_2,n_3,n_4}i_{n_1-1,n_2,n_3,n_4}\nn
	&&+c_{n_1,n_2,n_3,n_4-1}i_{n_1,n_2,n_3,n_4-1}+c_{n_1,n_2,n_3-1,n_4}i_{n_1,n_2,n_3-1,n_4}\nn
	&&+c_{n_1,n_2-1,n_3,n_4}i_{n_1,n_2-1,n_3,n_4}+c_{n_1-1,n_2,n_3,n_4}i_{n_1-1,n_2,n_3,n_4}\nn
	&&+c_{n_1,n_2,n_3,n_4}i_{n_1,n_2,n_3,n_4}+\delta_{4;r}=0~~~\label{recurbox}
	\eea
	Now we need to calculate the $\delta_{4}$ term.

	{\bf The Reduction of the boundary $\delta_{4}$ term:}
	Similarly to the former case, we could expand the $\delta_{4}$ term and taking the value of parameters $a_2$ to $a_{10}$ into the $\delta_{4}$ part. After this, we could get
	\bea
	\delta_{4;r}&=&\delta_{n_1+1,0}Q_{11;r}i_{-1,n_2,n_3,n_4}+\delta_{n_1,0}Q_{12;r}i_{-1,n_2+1,n_3,n_4}+\delta_{n_1,0}Q_{13;r}i_{-1,n_2,n_3+1,n_4}+\delta_{n_1,0}Q_{14;r}i_{-1,n_2,n_3,n_4+1}\nn
	&&+\delta_{n_1,0}Q_{15;r}i_{-1,n_2,n_3,n_4}+\delta_{n_2,0}Q_{21;r}i_{n_1+1,-1,n_3,n_4}+\delta_{n_2+1,0}Q_{22;r}i_{n_1,-1,n_3,n_4}+\delta_{n_2,0}Q_{23;r}i_{n_1,-1,n_3+1,n_4}\nn
	&&+\delta_{n_2,0}Q_{24;r}i_{n_1,-1,n_3,n_4+1}+\delta_{n_2,0}Q_{25;r}i_{n_1,-1,n_3,n_4}+\delta_{n_3,0}Q_{31;r}i_{n_1+1,n_2,-1,n_4}+\delta_{n_3,0}Q_{32;r}i_{n_1,n_2+1,-1,n_4}\nn
	&&+\delta_{n_3+1,0}Q_{33;r}i_{n_1,n_2,-1,n_4}+\delta_{n_3,0}Q_{34;r}i_{n_1,n_2,-1,n_4+1}+\delta_{n_3,0}Q_{35;r}i_{n_1,n_2,-1,n_4}+\delta_{n_4,0}Q_{41;r}i_{n_1+1,n_2,n_3,-1}\nn
	&&+\delta_{n_4,0}Q_{42;r}i_{n_1,n_2+1,n_3,-1}+\delta_{n_4,0}Q_{43;r}i_{n_1,n_2,n_3+1,-1}+\delta_{n_4+1,0}Q_{44;r}i_{n_1,n_2,n_3,-1}+\delta_{n_4,0}Q_{45;r}i_{n_1,n_2,n_3,-1}\nn
	~~~~\label{Wang-box-8}%
	\eea
	where the subscript "$r$" means the value of the parameter $Q$ after we set $a_2$ to $a_{10}$.
	%
	
	\subsubsection{The example: $I_4(1,1,1,2)$}
	Now we could use the recurrence relation \eref{recurbox} to calculate our example $I_4(1,1,1,2)$. Let $n_1=n_2=n_3=n_4=0$,we got (the coefficients of the other terms are all zero)
	\bea
	c_{0,0,0,0}i_{0,0,0,0}+c_{0,0,0,1}i_{0,0,0,1}+\delta_{4;0000}&=&0~~~~\label{Wang-box-9}
	\eea
	where $\delta_{4;0000}\equiv \delta_{4;r}|_{n_1=n_2=n_3=n_4=0}$. Translating to $I$, 
	we have the result
	\bea
	I_4(1,1,1,2)&=&c_{4\to1111}I_4(1,1,1,1)\nn
	&&+c_{4\to1110}I_4(1,1,1,0)+c_{4\to1101}I_4(1,1,0,1)+c_{4\to1011}I_4(1,0,1,1)+c_{4\to0111}I_4(0,1,1,1)\nn
	&&+c_{4\to2110}I_4(2,1,1,0)+c_{4\to2101}I_4(2,1,0,1)+c_{4\to2011}I_4(2,0,1,1)\nn
	&&+c_{4\to1210}I_4(1,2,1,0)+c_{4\to1201}I_4(1,2,0,1)+c_{4\to0211}I_4(0,2,1,1)\nn
	&&+c_{4\to1120}I_4(1,1,2,0)+c_{4\to1021}I_4(1,0,2,1)+c_{4\to0121}I_4(0,1,2,1)\nn
	&&+c_{4\to1102}I_4(1,1,0,2)+c_{4\to1012}I_4(1,0,1,2)+c_{4\to0112}I_4(0,1,1,2)~~~\label{box1112}
	\eea
	%
	with the coefficients
	\bea
	c_{4\to1111}&=&\frac{c_{0,0,0,0}}{c_{0,0,0,1}}(D-5)=\frac{Tr \hat Q_{ij;r}+(D-5)Q_{55;r}-\frac{D}{2}}{Q_{54;r}}\nn
	c_{4\to0111}&=&-\frac{Q_{15;r}\Gamma(D-3)}{c_{0,0,0,1}\Gamma(D-5)},~~
	c_{4\to1011}=-\frac{Q_{25;r}\Gamma(D-3)}{c_{0,0,0,1}\Gamma(D-5)}\nn
	c_{4\to1101}&=&-\frac{Q_{35;r}\Gamma(D-3)}{c_{0,0,0,1}\Gamma(D-5)},~~
	c_{4\to1110}=-\frac{Q_{45;r}\Gamma(D-3)}{c_{0,0,0,1}\Gamma(D-5)}\nn
	c_{4\to0211}&=&\frac{Q_{12;r}\Gamma(D-4)}{c_{0,0,0,1}\Gamma(D-5)},~~
	c_{4\to0121}=\frac{Q_{13;r}\Gamma(D-4)}{c_{0,0,0,1}\Gamma(D-5)},~~
	c_{4\to0112}=\frac{Q_{14;r}\Gamma(D-4)}{c_{0,0,0,1}\Gamma(D-5)}\nn
	c_{4\to2011}&=&\frac{Q_{21;r}\Gamma(D-4)}{c_{0,0,0,1}\Gamma(D-5)},~~
	c_{4\to1021}=\frac{Q_{23;r}\Gamma(D-4)}{c_{0,0,0,1}\Gamma(D-5)},~~
	c_{4\to1012}=\frac{Q_{24;r}\Gamma(D-4)}{c_{0,0,0,1}\Gamma(D-5)}\nn
	c_{4\to2101}&=&\frac{Q_{31;r}\Gamma(D-4)}{c_{0,0,0,1}\Gamma(D-5)},~~c_{4\to1201}=\frac{Q_{32;r}\Gamma(D-4)}{c_{0,0,0,1}\Gamma(D-5)},~~c_{4\to1102}=\frac{Q_{34;r}\Gamma(D-4)}{c_{0,0,0,1}\Gamma(D-5)}\nn
	c_{4\to2110}&=&\frac{Q_{41;r}\Gamma(D-4)}{c_{0,0,0,1}\Gamma(D-5)},~~c_{4\to1210}=\frac{Q_{42;r}\Gamma(D-4)}{c_{0,0,0,1}\Gamma(D-5)},~~c_{4\to1120}=\frac{Q_{43;r}\Gamma(D-4)}{c_{0,0,0,1}\Gamma(D-5)}~~~~\label{Wang-box-10}
	\eea
	Next  we need to use  the reduction of triangles with one double propagators given in \eref{Wang-tri-13}. 
	Put them into the \eref{box1112}, we got the complete reduction of box $I_4(1,1,1,2)$.
	\bea
	& & I_4(1,1,1,2)=c_{4\to4}I_4(1,1,1,1)\nn%
	&&+c_{4\to3;\bar1}I_4(0,1,1,1)+c_{4\to3;\bar2}I_4(1,0,1,1)+c_{4\to3;\bar3}I_4(1,1,0,1)+c_{4\to3;\bar4}I_4(1,1,1,0)\nn%
	&&+c_{4\to2;\bar1\bar2}I_4(0,0,1,1)+c_{4\to2;\bar1\bar3}I_4(0,1,0,1)+c_{4\to2;\bar1\bar4}I_4(0,1,1,0)\nn%
	&&+c_{4\to2;\bar2\bar3}I_4(1,0,0,1)+c_{4\to2;\bar2\bar4}I_4(1,0,1,0)+c_{4\to2;\bar3\bar4}I_4(1,1,0,0)\nn%
	&&+c_{4\to1;D_1}I_4(1,0,0,0)+c_{4\to1;D_2}I_4(0,1,0,0)+c_{4\to1;D_3}I_4(0,0,1,0)+c_{4\to1;D_4}I_4(0,0,0,1)~~~~\label{Wang-box-11}
	\eea
	with the long expressions of these coefficients  given in the companion Mathematica notebook. The result is confirmed by FIRE6. 
	
	%
	\subsection{The pentagon's case}
	The general form of pentagon is given by
	\bea
	I_5(n_1+1,n_2+1,n_3+1,n_4+1,n_5+1)&=&\int \frac{d^{D}l}{D_1^{n_1+1}D_2^{n_2+1}D_3^{n_3+1}D_4^{n_4+1}D_5^{n_5+1}}~~~~~\label{Wang-pen-1}
	\eea
	with
	\bea
	D_1&=&l^2-m_1^2,~~~~
	D_2=(l-p_1)^2-m_2^2,~~~~
	D_3=(l-p_1-p_2)^2-m_3^2\nn
	D_4&=&(l-p_1-p_2-p_3)^2-m_4^2,~~~
	D_5=(l+p_5)^2-m_5^2~~~~~\label{Wang-pen-2}
	\eea
	The parametric form of $I_5(n_1+1,n_2+1,n_3+1,n_4+1,n_5+1)$ could be written as
	\bea
	&&I_5(n_1+1,n_2+1,n_3+1,n_4+1,n_5+1)
	=\frac{i(-1)^{5+n_1+n_2+n_3+n_4+n_5}\Gamma(-\lambda_0)}{\sum_{i=1}^5 \Gamma(n_i+1)\Gamma(\lambda_6+1)}i_{\lambda_0;n_1,n_2,n_3,n_4,n_5}
	~~~~~\label{Wang-pen-3}
	\eea
	where
	\bea
	i_{\lambda_0;n_1,n_2,n_3,n_4,n_5}&=&\int d\Pi^{(6)} F^{\lambda_0}x_1^{n_1}x_2^{n_2}x_3^{n_3}x_4^{n_4}x_5^{n_5}x_6^{\lambda_6+1}\nn
	d\Pi^{(5)}&=&dx_1dx_2dx_3dx_4dx_5dx_6\delta(\sum x_j-1)\nn
	\lambda_0&=&-\frac{D}{2},~~~~
	\lambda_6=(D-6)-n_1-n_2-n_3-n_4-n_5~~~~~\label{Wang-pen-4}
	\eea
	and the function
	\bea
	U(x)&=&x_1+x_2+x_3+x_4+x_5\nn
	V(x)&=&x_1x_2p_1^2+x_1x_3p_{12}^2+x_1x_4p_{13}^2+x_1x_5p_{14}^2\nn
	&&+x_2x_3p_2^2+x_2x_4p_{23}^2+x_2x_5p_{24}^2+x_3x_4p_3^2+x_3x_5p_{34}^2+x_4x_5p_4^2\nn
	f(x)&=&(x_1+x_2+x_3+x_4+x_5)(m_1^2x_1+m_2^2x_2+m_3^2x_3+m_4^2x_4+m_5^2x_5)\nn
	&&-x_1x_2p_1^2-x_1x_3p_{12}^2-x_1x_4p_{13}^2-x_1x_5p_{14}^2-x_2x_3p_2^2-x_2x_4p_{23}^2-x_2x_5p_{24}^2\nn
	&&-x_3x_4p_3^2-x_3x_5p_{34}^2-x_4x_5p_4^2\nn
	F(x)&=&(x_1+x_2+x_3+x_4+x_5)(m_1^2x_1+m_2^2x_2+m_3^2x_3+m_4^2x_4+m_5^2x_5+x_6)\nn
	&&-x_1x_2p_1^2-x_1x_3p_{12}^2-x_1x_4p_{13}^2-x_1x_5p_{14}^2-x_2x_3p_2^2-x_2x_4p_{23}^2-x_2x_5p_{24}^2\nn
	&&-x_3x_4p_3^2-x_3x_5p_{34}^2-x_4x_5p_4^2~~~~~\label{Wang-pen-5}
	\eea
	where
	$p_{ij}\equiv p_i+p_{i+1}+\cdots p_{j-1}+p_{j}$.
	Now the matrix are given by
	\bea
	\hat A&=&\left[\begin{array}{cccccc}
		2m_1^2&m_1^2+m_2^2-p_1^2&m_1^2+m_3^2-p_{12}^2&m_1^2+m_4^2-p_{13}^2&m_1^2+m_5^2-p_1^2&1\nn
		m_1^2+m_2^2-p_1^2&2m_2^2&m_2^2+m_3^2-p_2^2&m_2^2+m_4^2-p_{23}^2&m_2^2+m_5^2-p_{24}^2&1\nn
		m_1^2+m_3^2-p_{12}^2&m_2^2+m_3^2-p_{23}^2&2m_3^2&m_3^2+m_4^2-p_3^2&m_3^2+m_5^2-p_{34}^2&1\nn
		m_1^2+m_4^2-p_{13}^2&m_2^2+m_4^2-p_{23}^2&m_3^2+m_4^2-p_3^2&2m_4^2&m_4^2+m_5^2-p_4^2&1\nn
		m_1^2+m_5^2-p_1^2&m_2^2+m_5^2-p_{24}^2&m_3^2+m_5^2-p_{34}^2&m_4^2+m_5^2-p_4^2&2m_5^2&1\nn
		1&1&1&1&1&0
	\end{array}\right]\nn
	~\hat K_A&=&\left[
	\begin{array}{cccccc}
		0&a_1&a_2&a_3&a_4&a_5\nn
		-a_1&0&a_6&a_7&a_8&a_9\nn
		-a_2&-a_6&0&a_{10}&a_{11}&a_{12}\nn
		-a_3&-a_7&-a_{10}&0&a_{13}&a_{14}\nn
		-a_4&-a_8&-a_{11}&-a_{13}&0&a_{15}\nn
		-a_5&-a_9&-a_{12}&-a_{14}&-a_{15}&0
	\end{array}
	\right]\nn~~~~~\label{Wang-pen-6}
	\eea
	Taking $B=-\frac{1}{x_6}$ and putting $z_i$ into the IBP identities
	\bea
	\sum_{i=1}^{6}\int \frac{\d}{\d x_i }\Big\{z_iF^{\lambda_0}x_1^{n_1}x_2^{n_2}x_3^{n_3}x_4^{n_4}x_5^{n_5}x_6^{\lambda_6+1}\Big\}+\delta_{5}&=&0~~~~~\label{Wang-pen-7}
	\eea
	where the $\delta_{5}$ is given by
	\bea
	\delta_{5}&=&\sum_{i=1}^{5}\delta_{\lambda_i,0}\int d\Pi^{(5)} \Big\{z_iF^{\lambda_0}x_1^{n_1}x_2^{n_2}x_3^{n_3}x_4^{n_4}x_5^{n_5}x_6^{\lambda_6+1}\Big\}|_{x_i=0}~~~\label{pen}
	\eea
	\subsubsection{Deriving the recurrence relation}
	Similarly to the previous subsections, expanding the IBP relation  we  get
	\bea
	&&\Big\{c_{n_1+1,n_2,n_3,n_4,n_5}1^{+}+c_{n_1+1,n_2-1,n_3,n_4,n_5}1^+2^-+c_{n_1+1,n_2,n_3-1,n_4,n_5}1^+3^-+c_{n_1+1,n_2,n_3,n_4-1,n_5}1^+4^-\nn
	&&+c_{n_1+1,n_2,n_3,n_4,n_5-1}1^+5^{-}+c_{n_1,n_2+1,n_3,n_4,n_5}2^+ +c_{n_1-1,n_2+1,n_3,n_4,n_5}1^-2^+ +c_{n_1,n_2+1,n_3-1,n_4,n_5}2^+3^-\nn
	&& +c_{n_1,n_2+1,n_3,n_4-1,n_5} 2^+ 4^- +c_{n_1,n_2+1,n_3,n_4,n_5-1} 2^+ 5^{-}+c_{n_1,n_2,n_3+1,n_4,n_5}3^+ +c_{n_1-1,n_2,n_3+1,n_4,n_5}1^- 3^+ \nn&&
	+c_{n_1,n_2-1,n_3+1,n_4,n_5}2^- 3^+ +c_{n_1,n_2,n_3+1,n_4-1,n_5}3^+4^- +c_{n_1,n_2,n_3+1,n_4,n_5-1}3^+5^-+c_{n_1,n_2,n_3,n_4+1,n_5}4^+ \nn
	&&+c_{n_1-1,n_2,n_3,n_4+1,n_5}1^-4^+ +c_{n_1,n_2-1,n_3,n_4+1,n_5}2^- 4^+ +c_{n_1,n_2,n_3-1,n_4+1,n_5} 3^{-}4^{+}+c_{n_1,n_2,n_3,n_4+1,n_5-1} 4^{+}5^{-}\nn
	&&+c_{n_1,n_2,n_3,n_4,n_5+1}5^{+}+c_{n_1-1,n_2,n_3,n_4,n_5+1}1^-5^+ +c_{n_1,n_2-1,n_3,n_4,n_5+1}2^-5^+ +c_{n_1,n_2,n_3-1,n_4,n_5+1}3^- 5^+ \nn
	&&+c_{n_1,n_2,n_3,n_4-1,n_5+1} 4^- 5^++c_{n_1-1,n_2,n_3,n_4,n_5}1^{-}+c_{n_1,n_2-1,n_3,n_4,n_5}2^{-}+c_{n_1,n_2,n_3-1,n_4,n_5}3^- \nn
	&&+c_{n_1,n_2,n_3,n_4-1,n_5} 4^- +c_{n_1,n_2,n_3,n_4,n_5-1}5^-+c_{n_1,n_2,n_3,n_4,n_5}\Big\}i_{n_1,n_2,n_3,n_4,n_5}+\delta_{5}=0~~~\label{912}
	\eea
	We could choose the particular value of parameter $a_2$ to $a_{15}$ to let the coefficients of the  first three line of \eref{912} be zero. The solution is 
	\bea
	a_2&=&\frac{-a_1}{\Delta_{pen}}|\tilde A_{13,56}|,~~a_3=\frac{a_1}{\Delta_{pen}}|\tilde A_{14,56}|,~~a_4=\frac{-a_1}{\Delta_{pen}}|\tilde A_{15,56}|,~~a_5=\frac{a_1}{\Delta_{pen}}|\tilde A_{16,56}|,~~a_6=\frac{a_1}{\Delta_{pen}}|\tilde A_{23,56}|\nn
	a_7&=&\frac{-a_1}{\Delta_{pen}}|\tilde A_{24,56}|,~~a_8=\frac{a_1}{\Delta_{pen}}|\tilde A_{25,56}|,~~a_9=\frac{-a_1}{\Delta_{pen}}|\tilde A_{26,56}|,~~a_{10}=\frac{a_1}{\Delta_{pen}}|\tilde A_{34,56}|,~~a_{11}=\frac{-a_{1}}{\Delta_{pen}}|\tilde A_{35,56}|\nn
	a_{12}&=&\frac{a_1}{\Delta_{pen}}|\tilde A_{36,56}|,~~a_{13}=\frac{a_1}{\Delta_{pen}}|\tilde A_{45,56}|,~~a_{14}=\frac{-a_1}{\Delta_{pen}}|\tilde A_{46,56}|,~~a_{15}=\frac{a_1}{\Delta_{pen}}|\tilde A_{56,56}|
	\eea
	where
	\bea
	\Delta_{pen}&=&\left|\begin{array}{cccc}
		A_{31}&A_{32}&A_{33}&A_{34}\nn
		A_{41}&A_{42}&A_{43}&A_{44}\nn
		A_{51}&A_{52}&A_{53}&A_{54}\nn
		A_{61}&A_{62}&A_{63}&A_{64}
	\end{array}\right|
	\eea
	After this, we got 
	\bea
	&&\Big\{c_{n_1,n_2,n_3,n_4,n_5+1;r}5^{+}+c_{n_1-1,n_2,n_3,n_4,n_5+1;r}1^-5^+ +c_{n_1,n_2-1,n_3,n_4,n_5+1;r}2^-5^+ +c_{n_1,n_2,n_3-1,n_4,n_5+1;r}3^- 5^+ \nn
	&&+c_{n_1,n_2,n_3,n_4-1,n_5+1;r} 4^- 5^+c_{n_1-1,n_2,n_3,n_4,n_5;r}1^{-}+c_{n_1,n_2-1,n_3,n_4,n_5;r}2^{-}+c_{n_1,n_2,n_3-1,n_4,n_5;r}3^- \nn
	&&+c_{n_1,n_2,n_3,n_4-1,n_5;r} 4^- +c_{n_1,n_2,n_3,n_4,n_5-1;r}5^-+c_{n_1,n_2,n_3,n_4,n_5;r}\Big\}i_{\lambda_0;n_1,n_2,n_3,n_4,n_5}+\delta_{5;r}=0~~~~~~~~~~\label{Wang-pen-8}
	\eea
	here we have defined
	\bea
	i^{+}i_{\lambda_0;n_1,n_2,n_3,n_4,n_5}&\equiv&i_{\lambda_0;n_1,\cdots n_i+1,\cdots n_5}\nn%
	i^{-}i_{\lambda_0;n_1,n_2,n_3,n_4,n_5}&\equiv &i_{\lambda_0;n_1,\cdots n_i-1,\cdots n_5}
	\eea
	with the coefficients
	\bea
	c_{0,0,0,0,1}&=&Q_{65;r}\lambda_6,~~c_{-1,0,0,0,1}=n_1Q_{15;r},~~c_{0,-1,0,0,1}=n_2Q_{25;r},~~c_{0,0,-1,0,1}=n_3Q_{35;r},~~c_{0,0,0,-1,1}=n_4Q_{45;r}\nn
	c_{-1,0,0,0,0}&=&n_1Q_{16;r},~~c_{0,-1,0,0,0}=n_2Q_{26;r},~~c_{0,0,-1,0,0}=n_3Q_{36;r},~~c_{0,0,0,-1,0}=n_4Q_{46;r},~~c_{0,0,0,0,-1}=n_5Q_{56;r}\nn
	c_{00000;r}&=& Tr \hat Q_{ij;r} +((D-6))Q_{66;rr}-\frac{D}{2}+n_1Q_{11;r}+n_2Q_{22;r}+n_3Q_{33;r}+n_4Q_{44;r}+n_5Q_{55;r}
	\eea
	while the matrix $\hat Q$ becomes to
	\bea
	\hat Q_{r}&=&\frac{1}{\Delta_{pen}}\left[\begin{array}{cccccc}
		\frac{1}{2}\Delta_{pen}&0&0&0&a_1|\tilde A_{1,6}|&a_1|\tilde A_{1,5}|\nn
		0&\frac{1}{2}\Delta_{pen}&0&0&-a_1|\tilde A_{2,6}|&-a_1|\tilde A_{2,5}|\nn
		0&0&\frac{1}{2}\Delta_{pen}&0&a_1|\tilde A_{3,6}|&a_1|\tilde A_{3,5}|\nn
		0&0&0&\frac{1}{2}\Delta_{pen}&-a_1|\tilde A_{4,6}|&-a_1|\tilde A_{4,5}|\nn
		0&0&0&0&\frac{1}{2}\Delta_{pen}+a_1|\tilde A_{5,6}|&a_1|\tilde A_{5,5}|\nn
		0&0&0&0&-a_1|\tilde A_{6,6}|&\frac{1}{2}\Delta_{pen}-a_1|\tilde A_{6,5}|
	\end{array}\right]
	\eea
	\subsection{Reducing the $\delta_{5}$ term}
	Similar to the former situation, the $\delta_{6;r}$ term is given by
	\bea
	\delta_{5;r}&=&Q_{11;r}\delta_{n_1,-1}i_{-1,n_2,n_3,n_4,n_5}
	+Q_{12;r}\delta_{n_1,0}i_{-1,n_2+1,n_3,n_4,n_5}+Q_{13;r}\delta_{n_1,0}i_{-1,n_2,n_3+1,n_4,n_5}\nn
	&&+Q_{14;r}\delta_{n_1,0}i_{-1,n_2,n_3,n_4+1,n_5}
	+Q_{15;r}\delta_{n_1,0}i_{-1,n_2,n_3,n_4,n_5+1}+Q_{16;r}\delta_{n_1,0}i_{-1,n_2,n_3,n_4,n_5}\nn%
	&&+Q_{21;r}\delta_{n_2,0}i_{n_1+1,-1,n_3,n_4,n_5}+Q_{22;r}\delta_{n_2,-1}i_{n_1,-1,n_2,n_3,n_4,n_5}+Q_{23;r}\delta_{n_2,0}i_{n_1,-1,n_3+1,n_4,n_5}\nn
	&&+Q_{24;r}\delta_{n_2,0}i_{n_1,-1,n_3,n_4+1,n_5}+Q_{25;r}\delta_{n_2,0}i_{n_1,-1,n_3,n_4,n_5+1}+Q_{26;r}\delta_{n_2,0}i_{n_1,-1,n_3,n_4,n_5}\nn%
	&&+Q_{31;r}\delta_{n_3,0}i_{n_1+1,n_2,-1,n_4,n_5}Q_{32;r}\delta_{n_3,0}i_{n_1,n_2+1,-1,n_4,n_5}+Q_{33;r}\delta_{n_3,-1}i_{n_1,n_2,-1,n_4,n_5}\nn
	&&+Q_{34;r}\delta_{n_3,0}i_{n_1,n_2,-1,n_4+1,n_5}+Q_{35;r}\delta_{n_3,0}i_{n_1,n_2,-1,n_4,n_5+1}+Q_{36;r}\delta_{n_3,0}i_{n_1,n_2,-1,n_4,n_5}\nn
	&&Q_{41;r}\delta_{n_4,0}i_{n_1+1,n_2,n_3,-1,n_5}+Q_{42;r}\delta_{n_4,0}i_{n_1,n_2+1,n_3,-1,n_5}+Q_{43;r}\delta_{n_4,0}i_{n_1,n_2,n_3+1,-1,n_5}\nn
	&&+Q_{44;r}\delta_{n_4,-1}i_{n_1,n_2,n_3,-1,n_5}+Q_{45;r}\delta_{n_4,0}i_{n_1,n_2,n_3,-1,n_5+1}+Q_{46;r}\delta_{n_4,0}i_{n_1,n_2,n_3,-1,n_5}\nn
	&&+Q_{51;r}\delta_{n_5,0}i_{n_1+1,n_2,n_3,n_4,-1}+Q_{52;r}\delta_{n_5,0}i_{n_1,n_2+1,n_3,n_4,-1}+Q_{53;r}\delta_{n_5,0}i_{n_1,n_2,n_3+1,n_4,-1}\nn
	&&+Q_{54;r}\delta_{n_5,0}i_{n_1,n_2,n_3,n_4+1,-1}+Q_{55;r}\delta_{n_5,-1}i_{n_1,n_2,n_3,n_4,-1}+Q_{56;r}\delta_{n_5,0}i_{n_1,n_2,n_3,n_4,n_5}
	\eea
	\subsection{The example: $I_5(1,1,1,1,2)$}
	Setting $n_1=n_2=n_3=n_4=n_5=0$, we got the IBP recurrence relation (other coefficients are all zero)
	\bea
	&&c_{0,0,0,0,1}i_{\lambda_0;0,0,0,0,1}+c_{0,0,0,0,0}i_{\lambda_0;0,0,0,0,0}+\delta_{5;00000}=0\nn
	\eea
	where $\delta_{5;00000}\equiv \delta_{5;r}|_{n_1=n_2=n_3=n_4=n_5=0}$.\\
	Comparing them with our scalar basis, we have the result
	\bea
	I_5(1,1,1,1,2)&=&c_{5\to5}I_5(1,1,1,1,1)+c_{5\to01111}I_4(0,1,1,1,1)+c_{5\to10111}I_5(1,0,1,1,1)\nn
	&&+c_{5\to11011}I_5(1,1,0,1,1)+c_{5\to11101}I_5(1,1,1,0,1)+c_{5\to11110}I_5(1,1,1,1,0)\nn
	&&+c_{5\to20111}I_5(2,0,1,1,1)+c_{5\to21011}I_5(2,1,0,1,1)+c_{5\to21101}I_5(2,1,1,0,1)\nn
	&&+c_{5\to21110}I_5(2,1,1,1,0)+c_{5\to02111}I_5(0,2,1,1,1)+c_{5\to12011}I_5(1,2,0,1,1)\nn
	&&+c_{5\to12101}I_5(1,2,1,0,1)+c_{5\to12110}I_5(1,2,1,1,0)+c_{5\to01211}I_5(0,1,2,1,1)\nn
	&&+c_{5\to10211}I_5(1,0,2,1,1)+c_{5\to11201}I_5(1,1,2,0,1)+c_{5\to11210}I_5(1,1,2,1,0)\nn
	&&+c_{5\to01121}I_5(0,1,1,2,1)+c_{5\to10121}I_5(1,0,1,2,1)+c_{5\to11021}I_5(1,1,0,2,1)\nn
	&&+c_{5\to11120}I_5(1,1,1,2,0)+c_{5\to01112}I_5(0,1,1,1,2)+c_{5\to10112}I_5(1,0,1,1,2)\nn
	&&+c_{5\to11012}I_5(1,1,0,1,2)+c_{5\to11102}I_5(1,1,1,0,2)
	\eea
	with the coefficients
	\bea
	c_{5\to5}&=&\frac{(D-6)c_{0,0,0,0,0}}{c_{0,0,0,0,1}},~~
	c_{5\to01111}=\frac{(D-6)(5-D)Q_{16;r}}{c_{0,0,0,0,1}},~~
	c_{5\to4;10111}=\frac{(D-6)(5-D)Q_{26;r}}{c_{0,0,0,0,1}}\nn
	c_{5\to4;11011}&=&\frac{(D-6)(5-D)Q_{36;r}}{c_{0,0,0,0,1}},~~c_{5\to4;11101}=\frac{(D-6)(5-D)Q_{46;r}}{c_{0,0,0,0,1}},~~c_{5\to4;11110}=\frac{(D-6)(5-D)Q_{56;r}}{c_{0,0,0,0,1}}\nn
	c_{5\to20111}&=&\frac{(D-6)Q_{21;r}}{c_{0,0,0,0,1}},~~c_{5\to21011}=\frac{(D-6)Q_{31;r}}{c_{0,0,0,0,1}},~~c_{5\to21101}=\frac{(D-6)Q_{41;r}}{c_{0,0,0,0,1}},~~c_{5\to21110}=\frac{(D-6)Q_{51;r}}{c_{0,0,0,0,1}}\nn
	c_{5\to02111}&=&\frac{(D-6)Q_{12;r}}{c_{0,0,0,0,1}},~~c_{5\to12011}=\frac{(D-6)Q_{32;r}}{c_{0,0,0,0,1}},~~c_{5\to12101}=\frac{(D-6)Q_{42;r}}{c_{0,0,0,0,1}},~~c_{5\to12110}=\frac{(D-6)Q_{52;r}}{c_{0,0,0,0,1}}\nn
	c_{5\to01211}&=&\frac{(D-6)Q_{13;r}}{c_{0,0,0,0,1}},~~c_{5\to10211}=\frac{(D-6)Q_{23;r}}{c_{0,0,0,0,1}},~~c_{5\to11201}=\frac{(D-6)Q_{43;r}}{c_{0,0,0,0,1}},~~c_{5\to11210}=\frac{(D-6)Q_{53;r}}{c_{0,0,0,0,1}}\nn
	c_{5\to01121}&=&\frac{(D-6)Q_{14;r}}{c_{0,0,0,0,1}},~~c_{5\to10121}=\frac{(D-6)Q_{24;r}}{c_{0,0,0,0,1}},~~c_{5\to11021}=\frac{(D-6)Q_{34;r}}{c_{0,0,0,0,1}},~~c_{5\to11120}=\frac{(D-6)Q_{54;r}}{c_{0,0,0,0,1}}\nn
	c_{5\to01112}&=&\frac{(D-6)Q_{15;r}}{c_{0,0,0,0,1}},~~c_{5\to10112}=\frac{(D-6)Q_{25;r}}{c_{0,0,0,0,1}},~~c_{5\to11012}=\frac{(D-6)Q_{35;r}}{c_{0,0,0,0,1}},~~c_{5\to11102}=\frac{(D-6)Q_{45;r}}{c_{0,0,0,0,1}}\nn
	\eea
	The final step is to reduce the coefficients of the general boxes to the scalar basis.

	After reduce them to our scalar basis, we got the final answer.
	\bea
	&&I_{5}(1,1,1,1,2)\nn
	&=&c_{5\to5}I_5(1,1,1,1,1)+c_{5\to4;\bar1}I_5(0,1,1,1,1)+c_{5\to4;\bar2}I_5(1,0,1,1,1)+c_{5\to4;\bar3}I_5(1,1,0,1,1)\nn
	&&+c_{5\to4;\bar4}I_5(1,1,1,0,1)+c_{5\to4;\bar5}I_5(1,1,1,1,0)+c_{5\to3;\bar1\bar2}I_5(0,0,1,1,1)+c_{5\to3;\bar1\bar3}I_5(0,1,0,1,1)\nn
	&&+c_{5\to3;\bar1\bar4}I_5(0,1,1,0,1)+c_{5\to3;\bar1\bar5}I_5(0,1,1,1,0)+c_{5\to3;\bar2\bar3}I_5(1,0,0,1,1)+c_{5\to3;\bar2\bar4}I_5(1,0,1,0,1)\nn
	&&+c_{5\to3;\bar2\bar5}I_5(1,0,1,1,0)c_{5\to3;\bar3\bar4}I_5(1,1,0,0,1)+c_{5\to3;\bar3\bar5}I_5(1,1,0,1,0)+c_{5\to3;\bar4\bar5}I_5(1,1,1,0,0)\nn
	&&+c_{5\to2;D_1D_2}I_5(1,1,0,0,0)+c_{5\to2;D_1D_3}I_5(1,0,1,0,0)+c_{5\to2;D_1D_4}I_5(1,0,0,1,0)+c_{5\to2;D_1D_5}I_5(1,0,0,0,1)\nn
	&&+c_{5\to2;D_2D_3}I_5(0,1,1,0,0)+c_{5\to2;D_2D_4}I_5(0,1,0,1,0)+c_{5\to2;D_2D_5}I_5(0,1,0,0,1)+c_{5\to2;D_3D_4}I_5(0,0,1,1,0)\nn
	&&+c_{5\to2;D_3D_5}I_5(0,0,1,0,1)+c_{5\to2;D_4D_5}I_5(0,0,0,1,1)+c_{5\to1;D_1}I_5(1,0,0,0,0)+c_{5\to1;D_2}I_5(0,1,0,0,0)\nn
	&&+c_{5\to1;D_3}I_5(0,0,1,0,0)+c_{5\to1;D_4}I_5(0,0,0,1,0)+c_{5\to1;D_5}I_5(0,0,0,0,1)
	\eea
	with the coefficients given in the attached Mathematica notebook.
	Now all coefficients are complete.
	\section{Analytic result of the coefficients}

	We give our analytic results  in Mathematica notebooks which are put on the  publicly available website at \href{https://github.com/Wanghongbin123/oneloop_parametric}{\url{https://github.com/Wanghongbin123/oneloop_parametric}}.
	\section{Summary and further discussion}
	In this paper, we consider the one-loop scalar integrals in the parametric representation given by Chen. However, in the recurrence relation, there are usually some terms that we do not want, as well as some terms with dimensional shifting in general, which makes our calculation not easy and efficient. In Chen's later paper \cite{chen2}, he used a method based on non-commutative algebra to cancel the dimension shift. Different from others methods, in the one loop case, we have used a straight method by solving the linear equation systems to simplify the IBP recurrence relation in the parametric representation. Benefited from the fact that the  $F$ is a homogeneous function of $x_i$ with degree two in one-loop's situation, we could solve the $x_i$ by $\frac{\d F}{\d x_i}$ with some free parameters. Then combining all the IBP identities with particular coefficients $z_i$, and then choose particular values of the free parameters, we  succeed to cancel the dimension shift and the terms with higher total power. As the complement of the tadpole coefficients in the reduction to our previous paper, we calculated several examples and gave the analytic result of the reduction.
	
	For further research, there are some questions needed to be considered. In the previous calculation, we can see that the coefficients we constructed $z_i$ is not polynomial since it has the denominator with the form $x_{n+1}^{\c}$, so we could not directly use the technic of syzygy. Also, the application of Chen's method to higher loop is definitely 
	another future direction. For this case the homogeneous function $F(x)$ is of degree $L+1$, where $L$ is the number of loops. For high loop's case, we should consider how to construct the coefficients $z_i$ efficiently, and find a relation similar to \eref{KA} to cancel the terms we do not need. Thirdly, the sub-topologies are totally decided by the boundary term in the parametric representation, and this may lead to some simplification of calculation.
	%
	\section*{Acknowledgments}
	I'd like to thank Bo Feng for the inspiring discussion and guidance. This work is supported by  Chinese NSF
	funding under Grant  No.11935013.
	\bibliographystyle{JHEP}
	\bibliography{reference2}
	\nocite{smir}
\end{document}